\documentclass{article}

\usepackage{arxiv}

\usepackage[utf8]{inputenc} 
\usepackage[T1]{fontenc}    
\usepackage[hidelinks]{hyperref}       
\usepackage{url}            
\usepackage{booktabs}       
\usepackage{amsfonts}       
\usepackage{nicefrac}       
\usepackage{microtype}      
\usepackage{lipsum}
\usepackage{graphicx}

\usepackage{amsmath,mathrsfs,enumerate,multirow,appendix}
\usepackage{longtable, caption}
\usepackage{xcolor, url}
\usepackage{natbib}

\usepackage[]{graphicx}
\chardef\bslash=`\\ 

\usepackage{tikz}

\usepackage{tikz-cd}
\tikzset{smalltext/.style={"\textup{\small #1}" description}}

\usetikzlibrary{shapes}
\usetikzlibrary{backgrounds}      
\tikzcdset{
  my CD/.style={                  
    arrow style=tikz,
    >={Triangle[length=2mm]},
    cells={nodes={inner sep=2mm}},
    row sep=1.5cm,  column sep=0.8cm}}
\tikzset{
  boxed/.style={                  
    show background rectangle,    
    background rectangle/.append style={ 
      draw=gray, thick, rounded corners}}}

\usepackage[superscript]{cite}

\makeatletter
\renewcommand\@biblabel[1]{#1.}
\makeatother

\graphicspath{ {./images/} }

\title{Generalizing the Finkelstein–Schoenfeld Test to Incorporate Multiple Alternating Thresholds}

\author{
 Yunhan Mou \\
  Department of Biostatistics,\\
  Yale School of Public Health\\
  New Haven, Connecticut, USA \\
  \texttt{yunhan.mou@yale.edu} \\
   \And
 Tassos Kyriakides \\
  Department of Biostatistics,\\
  Yale School of Public Health\\
  New Haven, Connecticut, USA; \\
  CSP Coordinating Center (CSPCC)\\
  VA Connecticut Health System\\
  West Haven, Connecticut, USA\\
  \texttt{tassos.kyriakides@yale.edu} \\
  \And
 Scott Hummel \\
  Frankel Cardiovascular Center,\\
  University of Michigan\\
  Ann Arbor, Michigan, USA;\\
  Department of Internal Medicine, Division of Cardiovascular Medicine,\\
  VA Ann Arbor Health System\\
  Ann Arbor, Michigan, USA\\
  \texttt{scothumm@umich.edu} \\
  \And
 Fan Li$^*$\\
    Department of Biostatistics,\\
    Center for Methods in Implementation and Prevention Science,\\
    Yale Center for Analytical Sciences,\\
    Yale School of Public Health\\
    New Haven, Connecticut, USA \\
    \texttt{fan.f.li@yale.edu}\\
  \And
  Yuan Huang\thanks{Corresponding authors}\\
    Department of Biostatistics,\\
    Yale Center for Analytical Sciences,\\
    Yale School of Public Health\\
    New Haven, Connecticut, USA; \\
    VA Connecticut Health System\\
    West Haven, Connecticut, USA\\
    \texttt{yuan.huang@yale.edu}\\
}

\begin{document}

\maketitle
\vspace{-0.1in}
\begin{abstract}
Composite endpoints consisting of both terminal and non-terminal events, such as death and hospitalization, are frequently used in cardiovascular clinical trials. The Finkelstein-Schoenfeld (FS) test provides a way to employ a hierarchical structure to combine fatal and non-fatal events by giving death information an absolute priority, which may limit the contribution of clinically meaningful non-fatal events. To provide a more flexible alternative, we propose the Finkelstein–Schoenfeld with Multiple Thresholds (FS-MT) test, which extends the standard FS test by incorporating multiple thresholds applied sequentially and alternating across endpoints. A weighted adaptive approach is also developed to help determine the thresholds in FS-MT. The proposed approach retains the statistical properties of the FS test while allowing more flexible use of information from lower-priority events. We evaluate the operating characteristics of the proposed test through simulations that vary the follow-up time, the correlation between events, and the treatment effect sizes. A case study based on the Digitalis Investigation Group clinical trial data is presented to further illustrate our proposed method. An R package ``FSMT'' that implements the proposed methodology has been developed.
\end{abstract}

\keywords{Clinical trials \and Composite Survival Endpoints \and Endpoint Hierarchy \and Generalized Pairwise Comparisons \and Win ratio \and Win statistics}

\section{Introduction} \label{intro}
\subsection{Background and literature review}
Composite time-to-event endpoints are frequently employed in cardiovascular and other clinical trials that evaluate the impact of new candidate treatments, to possibly improve statistical efficiency over a single endpoint and to dispense the need for multiplicity adjustment with multiple endpoints \citep{mao2021statistical}. For example, in a standard scenario, two common events used to define the composite endpoint are survival time and time to hospitalization. Such choices arise naturally since cardiovascular diseases have high morbidity and are more prevalent in elderly populations, where death is an immediate concern and a likely cause for informative censoring. The traditional analysis focuses on the time to the first event (either death or hospitalization). However, this approach often fails to consider the differential clinical importance between death and hospitalization, and does not have the capacity to address subsequent occurrence of a second component endpoint (e.g., a death event after a hospitalization event). 

The FS test originally proposed by \citet{finkelstein1999combining} offers an approach to address these potential limitations. This test relies on pairwise comparisons between participants' outcomes using a hierarchical structure, in which the survival time endpoint is often prioritized over non-fatal events. This procedure aligns the analysis with clinical priorities, ensuring that more serious events receive higher attention in evaluating treatment effects. In addition, as discussed by \citet{finkelstein1999combining}, another advantage of prioritized endpoints is to offer a systematic approach to combine the event time outcome with other types of clinical outcomes, such as longitudinal biomarkers, particularly when event outcomes necessitate a longer time to accumulate.
This concept of hierarchical pairwise comparisons was later formalized into the net benefit (NB) treatment effect measure by \citet{buyse2010generalized}, and was subsequently popularized in the cardiovascular research community through the introduction of the win ratio (WR) measure by \citet{pocock2012win}. In recent years, the WR and the broader framework of Generalized Pairwise Comparisons (GPC) have gained substantial attention. Owing to their advantages, an increasing number of clinical trials have adopted hierarchical pairwise comparison methods for analyzing composite endpoints \citep{gasparyan2021adjusted}. For instance, the EMPULSE trial (ClinicalTrials.gov identifier: NCT0415775), the VIP-ACS trial (NCT04001504), and the DAPA-HF trial (NCT03036124) included WR as the primary outcome measure in their protocols. Notably, the ATTR-ACT trial (NCT01994889), which evaluated Tafamidis for transthyretin amyloid cardiomyopathy, employed the FS test as its primary hypothesis testing method, resulting in the first U.S. Food and Drug Administration (FDA) approval of a therapy based on a GPC framework in 2019.

As the application of the GPC framework continues to expand in practical settings, there has been a growing focus on methodological research for hierarchical pairwise comparison methods. 
As an example, methods for statistical inference (e.g, in constructing confidence intervals and conducting hypothesis testing) with the WR measure were extensively studied \citep{luo2015alternative,bebu2016large,dong2016generalized,mao2019alternative}. The stratified WR was introduced to adjust for discrete baseline variables \citep{dong2018stratified, gasparyan2021adjusted}.
Beyond the FS test, NB and WR measures, the GPC framework now encompasses a rich family of related effect measures and methodological variations, including but not limited to the win–loss statistic \citep{luo2017weighted}, win probability \citep{gasparyan2021adjusted}, win odds \citep{brunner2021win}, and event-specific win ratio \citep{yang2021eventct}. Each of these measures shares a similar pairwise comparison foundation but offers different ways to summarize or weight the comparative information. 
There have also been methodological developments for the GPC framework in general, such as statistical inference methods (see \citet{ozenne2025inference} for a review and comparison), stratification \citep{dong2023stratified, backer2025stratification}, handling of censoring \citep{peron2018extension, deltuvaite2023generalized, deltuvaite-thomas2025rightcensored}, and beyond. See \citet{dong2023win}, \citet{verbeeck2023generalized}, and \citet{buyse2025handbook} for further details and discussions on the GPC framework.

\subsection{Objectives and contributions}
Despite its successful usage and growing popularity, testing the null hypothesis with the standard FS test (or other approximately equivalent hypothesis testing procedures within the GPC framework) may sometimes limit sensitivity to treatment effects on lower-priority endpoints.
As discussed by \citet{redfors2020win}, when the top layer dominates comparisons, such as when mortality is frequent and largely unaffected by treatment, potential signals at subsequent layers may be underrepresented, which is an inherent feature of comparison schemes with a strict hierarchical structure.
To provide an alternative perspective under such settings, we propose the Finkelstein–Schoenfeld with Multiple Thresholds (FS-MT) test, which builds upon the FS framework while allowing more flexible use of information from lower-priority endpoints.
Specifically, our win function is based on pairwise comparisons that employ multiple pre-specified thresholds applied sequentially and alternating across endpoints, such that a definitive win (or loss) is declared only when the observed difference exceeds these thresholds. 
Ultimately, the thresholds are shrunk to minimal clinically meaningful differences.
This is in contrast to the definition of the standard win functions, for which a single threshold is typically defined for each endpoint.
While such minimal clinically meaningful differences may be nonzero, for example, a difference of 5 points in the Kansas City Cardiomyopathy Questionnaire Overall Score is often considered clinically significant \citep{voors2022sglt2}, we use zero in our illustration for simplicity. Nonzero thresholds are naturally supported by both the standard FS test and the proposed FS-MT test.

The general concept of looking at multiple thresholds in pairwise comparisons dates back to \citet{buyse2010generalized}. Specifically, for the NB measure, they considered multiple comparison thresholds, for which the successive thresholds were designed to reflect different levels of clinical differences. However, their initial work did not explicitly explore the use of multiple thresholds in settings involving more than one endpoint, particularly not in scenarios where thresholds are to be applied in an alternating manner across different endpoints.
In follow-up work \citep{peron2016assessment, peron2019benefit}, multiple thresholds were also considered separately in individual tests for sensitivity analysis. In general, the implications of using multiple thresholds for each endpoint within a single test have not been fully elucidated, nor there exist general operational guidelines when multiple thresholds for each endpoint are considered in a single test. Moreover, no prior work has integrated multiple thresholds and multiple outcomes within a unified testing framework; hence our proposed FS-MT method fills this gap. In our setting with multiple outcomes, the motivation for using multiple thresholds in FS-MT is to provide lower layers higher opportunities to contribute information to the win function when necessary. Finally, in addition to formalizing the architecture with multiple thresholds for multiple endpoints, we also introduce a data-driven weighted adaptive approach to automatically determine possible thresholds, offering some convenience in specifying the thresholds without affecting the test size.

To summarize, as a modified version of FS test, the FS-MT test maintains the statistical properties of the standard FS test but relaxes its strict hierarchical structure with intuitive modifications to the comparison process. At each layer, the FS-MT test retains all the ties in the FS test but could lead to different results without ties. 
To better demonstrate the performance of the FS-MT test, we compare the proposed FS-MT test with the FS test via simulations. In our simulations, three critical influencing factors, namely the maximal follow-up time, the correlation between two endpoints, and the treatment effect size, are considered. 
For further illustration, we apply both methods to reanalyze data from the Digitalis Investigation Group (DIG) clinical trial (NCT00000476). 

\section{Statistical Methods}\label{method}

In this section, we first review the standard FS test and commonly used treatment effect measures in the GPC framework. Then we describe our proposed FS-MT test under a simple yet common setting with two component time-to-event endpoints and two thresholds per endpoint, followed by a weighted adaptive approach to determine thresholds. A general formulation of FS-MT with multiple component endpoints and an arbitrary number of thresholds will be provided in Supporting Information S1. Starting with the simple setting, we assume there are two layers, where the first layer is survival time and the second layer is time to hospitalization, which is a typical non-fatal event of clinical importance. We assume a clinical trial with $n$ participants and two treatment arms with $Z_i=0$ if the participant is in the control group and $Z_i=1$ if in the treatment group. For the $i$-th participant, let $D_i$ be the (observed) survival time, $\Delta_{Di}$ denote the censoring indicator for $D_i$ such that $\Delta_{Di}=0$ if death event is observed and $\Delta_{Di}=1$ if the death time is right censored. We write $T_{1i}$ to denote the observed time to hospitalization, and similarly $\Delta_{1i}$ to denote the censoring indicator; that is $\Delta_{Ti}=0$ if non-fatal event is observed and $\Delta_{Ti}=1$ if right censored. Throughout, we let $I(\cdot)$ be the indicator function and $\text{Sign}(\cdot)$ be the sign function.

\subsection{Standard FS test: A review}\label{FS:review}

The FS test is based on pairwise comparisons across all study participants. For the comparison between participants $i$ and $j$, we define the win-loss score $U_{ij}$ to be 1 if $i$ has a more favorable outcome (win) than $j$, $-1$ if $i$ has a less favorable outcome (loss) than $j$, and 0 if neither is more favorable hence the comparison is uninformative or indeterminate (tie). Within each pairwise comparison, for the calculation of $U_{ij}$, one first examines the survival time to assess if one participant lives longer than the other. If participants $i$ and $j$ have the same days alive, or the comparison is uninformative due to censoring (hence tied on survival time), the time to hospitalization is then compared to assess if one participant has experienced a longer time before hospitalization than the other. If the time to hospitalization is the same or uninformative, a tie will be concluded for the comparison between participants $i$ and $j$. 

To test the null hypothesis of no treatment effect with such composite endpoints, the FS test considers the following test statistic:
$$
S=\sum_{i=1}^{n} Z_i \sum_{j=1}^{n}U_{ij}.
$$
Here, $S$ is the sum of total win-loss scores (net win scores), obtained by comparing each treated participant against all participants. \citet{finkelstein1999combining} have shown that, under the null, $S$ follows an asymptotic normal distribution with mean zero and admits a closed-form variance estimator given by 
$$
\widehat{\text{Var}}(S) = \frac{\sum_{i=1}^{n} Z_i\left(n-\sum_{i=1}^{n} Z_i\right)}{n(n-1)}\left( \sum_{i=1}^n \left(\sum_{j=1}^{n}U_{ij}\right)^2\right).
$$

The FS test can be viewed as a hypothesis testing method within the GPC framework \citep{buyse2010generalized,buyse2025handbook}. While the FS test focuses on hypothesis testing, the pairwise comparison results in the GPC framework can also be used to construct treatment effect measures that quantify both the magnitude and direction of treatment benefit for composite endpoints. Let $N_+=\sum_{i=1}^{n} Z_i \sum_{j=1}^{n} (1-Z_j) I(U_{ij}=1)$, $N_-=\sum_{i=1}^{n} Z_i \sum_{j=1}^{n} (1-Z_j) I(U_{ij}=-1)$, $N_0=\sum_{i=1}^{n} Z_i \sum_{j=1}^{n} (1-Z_j) I(U_{ij}=0)$ denote the total numbers of wins, losses, and ties obtained by treatment participants when compared against control participants, respectively. It is immediate that the FS test statistic is given by $S=(N_+-N_-)+\sum_{i=1}^{n} \sum_{j=1}^{n} Z_i Z_jU_{ij}$. Based on these quantities, treatment effect measures commonly used in the GPC framework can be calculated. For example, net benefit (NB), win odds (WO), and win ratio (WR) are obtained as:
\begin{equation} \label{GPC_measures}
\text{NB} = \frac{N_+ - N_-}{N_+ + N_- + N_0}, \qquad  
\text{WO} = \frac{N_+ + 0.5N_0}{N_- + 0.5N_0}, \qquad  
\text{WR} = \frac{N_+}{N_-}.
\end{equation}
For these three GPC measures, $\text{NB} = 0$, $\text{WO} = 1$, and $\text{WR} = 1$ indicate no treatment effect, whereas $\text{NB} < 0$, $\text{WO} < 1$, and $\text{WR} < 1$ suggest a potential negative treatment effect, and the opposite directions ($\text{NB} > 0$, $\text{WO} > 1$, $\text{WR} > 1$) imply a positive treatment effect.
These estimators are also referred to as win statistics, and methods for statistical inference have been developed based on each win statistic, leveraging the theory of U-statistics, randomization-based procedure, and beyond; see, for example, \citet{buyse2010generalized}, \citet{luo2015alternative}, \citet{bebu2016large}, \citet{brunner2021win}, and \citet{buyse2025handbook}. In what follows, we focus primarily on hypothesis testing by introducing a new variant of the FS test that accommodates alternating thresholds across endpoints.

\subsection{Alternating multiple thresholds across endpoints} \label{method_WR-MT}

In order to release the strict hierarchical structure of the FS test and allow for a controllable degree of priority that survival time has against time to hospitalization, we propose to alternate multiple thresholds across endpoints in the original FS test, termed FS-MT test. With the FS-MT test, one first compares two participants with larger thresholds through the hierarchical structure before moving to the set of comparisons based on lower thresholds. 
The idea of the FS-MT test is illustrated in Figure \ref{fig: flow char FS-MT}, where the first column of comparisons describes the comparisons under the larger thresholds $d$ and $t_1$, and the second column of comparisons describes the comparisons using 0 as the threshold. It is evident that the standard FS test is a special case of the FS-MT test, by only involving the second column of comparisons for calculating the win-loss score.

\begin{center}
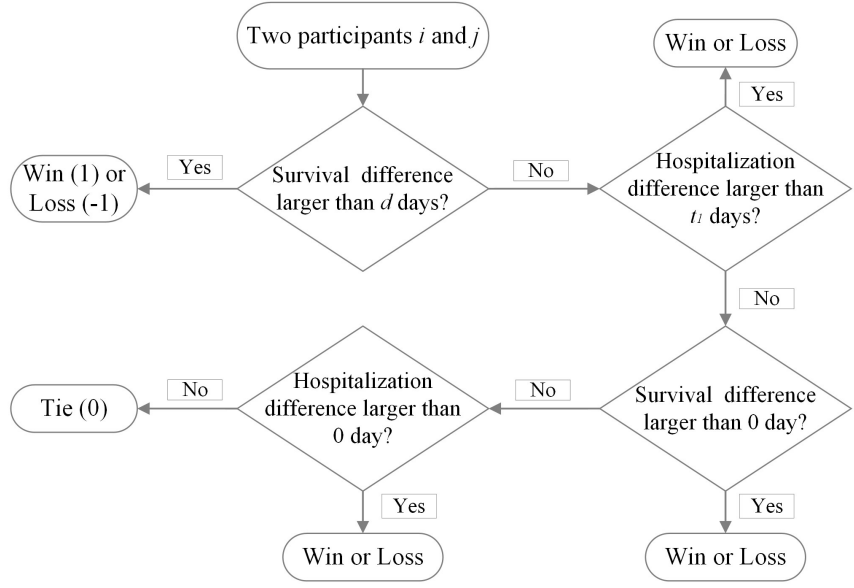

\begin{tikzcd}[cells={nodes={draw=black, rectangle, anchor=center}}, row sep=1.5cm, column sep=0.8cm]
  & \textup{\shortstack{Two participants $i$ and $j$}}\arrow[d, "Compare"] & & \\
  \textup{\shortstack{Win $(1)$ or \\ Loss $(-1)$}} & 
  \textup{\shortstack{Survival difference \\ larger than $d$ days?}}\arrow[l, "Yes"']\arrow[d, "No"] &  
  \textup{\shortstack{Survival difference \\ larger than $0$ days?}}\arrow[r, "Yes"]\arrow[d, "No"] & 
   \textup{\shortstack{Win $(1)$ or \\ Loss $(-1)$}}\\
   \textup{\shortstack{Win $(1)$ or \\ Loss $(-1)$}} & 
  \textup{\shortstack{Hospitalization \\ difference larger than \\ $d$ days?}}\arrow[l, "Yes"']\arrow[ru, "No"'{pos=0.275}, to path={
    -- ([xshift=1.8em]\tikztostart.east) 
    |- (\tikztotarget)\tikztonodes}]
  &  
  \textup{\shortstack{Hospitalization \\ difference larger than \\ $0$ days?}}\arrow[r, "Yes"]\arrow[d, "No"] &
   \textup{\shortstack{Win $(1)$ or \\ Loss $(-1)$}}\\
   & & \textup{\shortstack{Tie $(0)$}} &
\end{tikzcd}
\captionof{figure}{A schematic procedure of pairwise comparisons between two participants, $i$ and $j$, in the FS-MT test with thresholds $(d,0)$ for the survival time endpoint and $(t_1,0)$ for the time-to-hospitalization endpoint.}
\label{fig: flow char FS-MT}
\end{center}

To formally introduce our approach, we first define the comparison functions for survival time and time-to-hospitalization endpoints, under thresholds $d\geq 0$ and $t_1\geq 0$, denoted by $\mathscr{D} _{ij}(d)$ and $ \mathscr{T} _{ij1}(t_1)$, as follows: 
\begin{eqnarray*}
    \mathscr{D} _{ij}(d)  &=& I \left(|D_i-D_j|\geq d \right) \\
                          && \times 
                            \Big[  (1-\Delta_{Di})(1-\Delta_{Dj})\text{Sign}(D_i-D_j) 
                                    + \Delta_{Di}(1-\Delta_{Dj})I(D_i \geq D_j) \\
                                &&\quad - (1-\Delta_{Di})\Delta_{Dj}I(D_i \leq D_j)
                            \Big], \\
    \mathscr{T} _{ij1}(t_1) &=& I \left(|T_{1i}-T_{1j}|\geq t_1 \right) \\
                            && \times 
                            \Big[ (1-\Delta_{1i})(1-\Delta_{1j})\text{Sign}(T_{1i}-T_{1j})
                                    + \Delta_{1i}(1-\Delta_{1j})I(T_{1i} \geq T_{1j})\\
                                &&\quad - (1-\Delta_{1i})\Delta_{1j}I(T_{1i} \leq T_{1j})
                            \Big].
\end{eqnarray*}
Taking the first comparison function, $\mathscr{D} _{ij}(d)$, as an example, we examine the following scenarios. (1) When death events of both participants $i$ and $j$ are observed (i.e., $\Delta_{Di}=\Delta_{Dj}=0$), $\mathscr{D} _{ij}(d)=1$ represents a win for participant $i$ if $i$ lives at least $d$ days longer than $j$, or $\mathscr{D} _{ij}(d)=-1$ represents a loss if $j$ lives at least $d$ days longer than $i$; (2) When death event is observed for participant $j$ only (i.e., $\Delta_{Di}=1,\Delta_{Dj}=0$), $\mathscr{D} _{ij}(d)=1$ represents a win for the participant $i$ if the known survival time of $i$ is already at least $d$ days longer than $j$; (3) When death event is observed for participant $i$ only (i.e., $\Delta_{Di}=0,\Delta_{Dj}=1$), $\mathscr{D} _{ij}(d)=-1$ represents a loss for the participant $i$ if the known survival time of $j$ is already at least $d$ days longer than $i$; (4) Under other situations, $\mathscr{D} _{ij}(d)=0$ represents a tie for the comparison between them.

Using the comparison functions defined above, the process illustrated in Figure \ref{fig: flow char FS-MT}  can be described in the following four stages.
\begin{enumerate}
\item[] \hspace{-0.15in} Stage 1: Compare the survival difference at the $d$ level. That is:
    \begin{equation*}
        U_{ij1} = \mathscr{D}_{ij}(d).
    \end{equation*}

\item[] \hspace{-0.15in} Stage 2: If $U_{ij1}=0$, compare time-to-hospitalization difference at the $t_1$ level. That is:
    \begin{equation*}
        U_{ij2} = I(U_{ij1}=0)\times \mathscr{T}_{ij1}(t_1).
    \end{equation*}

\item[] \hspace{-0.15in} Stage 3: If $U_{ij2}=0$, compare any survival difference between $i$ and $j$. That is:
    \begin{equation*}
        U_{ij3} = I(U_{ij1}=U_{ij2}=0)\times \mathscr{D}_{ij}(0).
    \end{equation*}

\item[] \hspace{-0.15in} Stage 4: If $U_{ij3}=0$, compare any time-to-hospitalization difference between $i$ and $j$. That is:
    \begin{equation*}
        U_{ij4} = I(U_{ij1}=U_{ij2}=U_{ij3}=0)\times \mathscr{T}_{ij1}(0).
    \end{equation*}
    
\item[] \hspace{-0.15in} Summing the above four components, we obtain the win-loss score for the pair $i$ and $j$ as (including the set of thresholds in the argument for the win-loss score):
    \begin{equation*}
        U_{ij}(d,t_1,0,0) = U_{ij1} + U_{ij2} + U_{ij3} + U_{ij4}. 
    \end{equation*}
\end{enumerate}

Once the win-loss score is computed for each pair, hypothesis testing with the FS test can proceed following the general procedure outlined in Section \ref{FS:review}.

In general, the number of stages of a FS-MT test can be increased but should be pre-specified. For example, a 6-stage FS-MT test can be formed by specifying a sequence $\{d_1,t_{11},d_2,t_{12}, 0, 0\}$ and by requiring the comparison to go along the thresholds sequentially. To this end, the thresholds for the same information in later stages should always be smaller than the ones in higher stages (e.g., $d_2<d_1,t_{12}<t_{11}$) to make the added new stage non-trivial. In practice with two layers, we recommend a 4-stage FS-MT test, which usually provides sufficient control of the priorities between survival time and time-to-hospitalization endpoints. 
In Section \ref{se:inf of cor}, we will show in simulation that by employing the last two stages, which correspond exactly to those used in the standard FS test, the proposed method yields the same number of ties as the FS test.

In the 4-stage FS-MT test illustrated above, there are two opportunities for the time-to-hospitalization endpoint to contribute to the result of a pairwise comparison, as opposed to the single opportunity provided by the standard FS test. Such additional opportunity releases the strict hierarchy. As two extreme examples, when $d=+\infty, t_1=0$, the FS-MT test prioritizes the time-to-hospitalization endpoint strictly just like the standard FS test prioritizes the survival time endpoint. In contrast, when $d=0,t_1=0$, the FS-MT test reduces to standard FS test. Therefore, by varying the thresholds, one may control the priority between two endpoints from one end to the other. Such flexibility allows the convenient inclusion of clinical priors. 

\subsection{Adaptive thresholds}
\label{method_adaptive thresholds}

To improve the practical utility of the FS-MT testing procedure, in this section, we introduce a pre-specified data-driven adaptive approach to choose the thresholds. This can be useful in the absence of sufficient clinical information that determines the multiple thresholds.
To begin with, we define two sequences of all non-zero differences in pairwise comparisons for survival time and time to hospitalization accordingly:
\begin{eqnarray*}
   \mathbb{D} &=& \left\{ |D_i - D_j| : \forall i,j \in \{1,2,...,n\}, 
                    (D_i - D_j) \neq 0  \right\}, \\
   \mathbb{T}_1 &=& \left\{ |T_{1i} - T_{1j}| : \forall i,j \in \{1,2,...,n\}, 
                    (T_{1i} - T_{1j}) \neq 0 \right\}.
\end{eqnarray*}
These two sequences take comparisons between two participants from the same (i.e., both receive treatment or control) or contrasting groups (i.e., one receives treatment and the other receives control) into account, which is essential for gaining the adaptive property (will be discussed shortly). Then, with the calipers $c_1,c_2\in [0,1]$ specified, we calculate the empirical $c_1,c_2$ quantile values of these two sequences correspondingly, denoted as $q_D^{c_1}=\hat q_{c_1}(\mathbb{D})$ and $q_{1}^{c_2}=\hat q_{c_2}(\mathbb{T}_1)$. Next, we set $(d,t_1)=(q_D^{c_1},q_{1}^{c_2})$, where, loosely speaking, the relationship between $c_1$ and $c_2$ controls the weights of survival time and time-to-hospitalization endpoints. Intuitively, a larger caliper $c$ corresponds to a larger threshold, indicating that less emphasis is placed on an endpoint. Therefore, $c_1\leq c_2$ is usually adopted as we shall still prioritize the higher layer even though the strict hierarchy is intended to be released. In general, the influence of calipers on the weight between endpoints may be more intuitive than that of thresholds themselves since the thresholds rely on their relative scale with the observed differences.

To control the weight in a practical fashion, we consider employing $c_1=c_2=c$ and a weight parameter $w>0$: after obtaining $q_D^{c}=\hat q_{c}(\mathbb{D})$ and $q_{1}^{c}=\hat q_{c}(\mathbb{T}_1)$, we set $(d,t_1)=(q_D^{c},q_{1}^{c}/w)$, where the value of $w$, instead of the relationship between $c_1$ and $c_2$ controls the weight between two endpoints. A smaller $w$ indicates that less emphasis is placed on the time-to-hospitalization endpoint, as it increases the threshold for this endpoint. Since the remaining tuning parameters are $w,c$, we denote the resulted $\text{FS-MT}(q_D^{c},q_{1}^{c}/w,0,0)$ test as $\text{FS-AT}(w)^{c}$ test, which stands for the FS test with adaptive thresholds by weight $w$ and shared caliper $c$. In Supporting Information S1, we present the generalized formulas for adaptive thresholds under multiple non-fatal events for completeness.

In the process of constructing adaptive thresholds, we note that only non-zero differences are included in the sequences to exclude the results of self-comparison and ties. This ensures the subsequent quantiles are non-trivial (i.e., $d\neq0,t_1\neq0$). The caliper $c$ then indicates $100c\%$ smallest non-zero absolute differences shall be considered as ties in Stages 1 and 2. As an empirical rule of thumb, we recommend using $c=20\%$ as the default value when clinical priors are lacking, and its impact on the FS-AT test will be demonstrated in ensuing simulation study. When the default caliper value is employed, for simplicity, we omit the superscript $c=20\%$ in the above notations and denote $q_D=q_D^{20\%}=\hat q_{20\%}(\mathbb{D}),q_{1}=q_{1}^{20\%}=\hat q_{20\%}(\mathbb{T}_1), \text{FS-AT}(w)=\text{FS-AT}(w)^{20\%}$. As for the weight parameter $w$, a smaller $w$ assigns a smaller weight to the time-to-hospitalization endpoint by elevating the thresholds for differences in the time to hospitalization in Stage 2. Since the differential clinical importance between two endpoints is already addressed by the hierarchy and the caliper, the default value $w=1$ is often an adequate choice without external clinical knowledge. In addition, we will show that the FS-AT test is generally robust to the choice of several common values of $w$ through simulations.

Operationally, the adaptive thresholds in $\text{FS-AT}(w)^{20\%}$ test can be obtained as follows:
\begin{enumerate}
    \item[] \hspace{-0.15in} Step 1: Observe the survival time and time to hospitalization with censoring indicators for all participants, i.e., 
    $\{D_i, \Delta_{D_i}\}_{i=1}^n$ and $\{T_i,\Delta_{T_i}\}_{i=1}^n$;
    \item[] \hspace{-0.15in} Step 2: For each of the  $\{D_i, \Delta_{D_i}\}_{i=1}^n$ and $\{T_i,\Delta_{T_i}\}_{i=1}^n$, obtain pairwise absolute time differences and keep non-zero differences only, which form the sequence $\mathbb{D}$ and $\mathbb{T}_1$, respectively;
    \item[] \hspace{-0.15in} Step 3: Take the 20\% empirical quantile of the above two absolute time difference sequences, $q_D^{20\%}=\hat q_{20\%}(\mathbb{D})$
    and $q_{T_1}^{20\%}=\hat q_{20\%}(\mathbb{T}_1)$;
    \item[] \hspace{-0.15in} Step 4: Apply weight $w$ to obtain thresholds for survival time and time-to-hospitalization endpoints as $d=q_D^{20\%}$ and $t_1=q_{1}^{20\%}/w$, respectively.
\end{enumerate}

We provide several additional remarks before presenting the simulation study. First, the validity of the hypothesis test with the FS-MT test remains unaffected by the specification of thresholds. 
Under the null hypothesis of no treatment effect across endpoints, the choice of thresholds is not expected to introduce spurious signals or differential outcomes, thereby preserving the nominal type I error rate asymptotically. 
This property provides some justifications for using adaptive thresholds derived empirically from the observed data (see Table \ref{tab:sim s0} for empirical type I error rates for the FS-AT test). However, as elaborated below, the choice of thresholds can influence power. In this regard, threshold specification under the FS-MT test is reminiscent of test statistic specification in permutation tests, which has an impact on power without compromising the test validity.
Second, the use of adaptive thresholds leads to more adaptive hypothesis testing procedure (but still based on a pre-specified rule). For example, when the treatment only has an effect on extending the time to hospitalization, the FS-AT test improves power compared to the standard FS test. On the other hand, when the treatment only has an effect on survival, the FS-AT test maintains comparable power to the standard FS test. This adaptive property arises from estimating thresholds using pooled comparisons (i.e., pairwise comparisons among all participants, regardless of treatment group) and then applying these thresholds to comparisons between treatment arms. 
If no treatment effect exists on the survival time layer, the distribution of absolute survival differences between contrasting treatment groups resembles that of the pooled population. In this case, using the default $c = 20\%$, approximately 20\% of non-zero differences between treatment and control pairs will result in ties and move the comparison to the next endpoint (e.g., time to hospitalization). This reduces the influence of random variation on the survival layer.
Conversely, when a treatment effect exists on survival time, the distribution of absolute survival differences between treatment groups will shift rightward relative to the pooled distribution. As a result, the same 20\% quantile threshold now classifies fewer contrasting pairs as ties in Stage 1, thereby allowing more comparisons to be resolved at the survival layer. Taken together, these properties make the FS-AT test potentially appealing in practice, especially when specifying fixed thresholds a priori is challenging.

\section{Simulation Study}
\label{emp com}
In this section, we examine the performance of the proposed test empirically via simulation. We first compare the proposed test and standard FS test regarding empirical power (after validating their empirical type I error rates), and then investigate the influence that caliper $c$ and weight $w$ have on the empirical power. Although the proposed test can accommodate multiple nonfatal events, we focus on the simpler setting with two endpoints (survival time and time to hospitalization) for a demonstration.

\subsection{Simulation setup}\label{sim setup}
We consider a two-arm clinical trial with a total sample size $n=2000$ and equal allocation, where half of them are assigned to the treatment group $Z=1$ and the other half to the control group $Z=0$. We simulate the two correlated times to represent potentially latent survival time in days and time to hospitalization in days using a copula approach. Following \citet{luo2015alternative}, we adopt the Gumbel-Hougaard copula with a bivariate distribution that has exponential margins. We let $h_{D}(Z)=\lambda_D \exp(-\alpha_D Z)$ be the hazard rate for death event, and $h_{T_1}(Z)=\lambda_{T_1} \exp(-\alpha_{T_1} Z)$ be the hazard rate for hospitalization event. Then, the vector of survival time and time to hospitalization (in days) $(D^*,T_{1}^*)$ has the joint survival function: 
$$P(D^*>y_1, T_{1}^*>y_2|Z) = \exp\left\{-[(h_{D}(Z) y_1)^\beta + (h_{T_1}(Z) y_2)^\beta]^{(1/\beta)}\right\},$$ 
where $\beta \ge 1$ controls the correlation between the two endpoints (i.e., Kendall’s concordance: $1-1/\beta$). We fix parameters $\lambda_D=0.0008, \lambda_{T_1}=0.0022$ and let $\alpha_D,\alpha_{T_1}\in\{0,0.1,0.2,0.3\}$, in which $\{0,0.1,0.2,0.3\}$ stand for no, very weak, weak, and modest treatment effects. We assume treatment is not worse than control (in terms of extending survival time or time to hospitalization). In our simulation, it is possible to have sample $i$ with $T_{1i}^* \ge D^*_i$, which indicates there is no hospitalization event for $i$. Observed survival time and time to hospitalization are then obtained by performing administrative censoring after the scheduled days of follow-up for all participants, with the scheduled follow-up length denoted as $\text{FU}$. A nominal level of $\alpha=0.05$ for a two-sided test is employed throughout.

We compare the FS-AT test under the default weight and caliper choices (i.e., $\text{FS-AT}(w=1)^{c=20\%}$) with the standard FS test. In addition, we primarily evaluate the performance of the FS-AT test under various choices of weights and calipers, with one exception in assessing the influence of calipers, where a FS-MT test composed of eight stages is included for comparison.
Three key factors, follow-up time, correlation between endpoints, and treatment effect size, are varied and detailed in Table \ref{tab:sim sce} for simulation scenarios S1-S8. Additional simulation scenarios are explored in Supporting Information S2. The empirical power is based on 2000 simulation replicates, and the empirical type I error rate calculation is based on 5000 replicates.

All computations are implemented in R, and include function calls from \texttt{gumbel} \citep{gumbel}, \texttt{doParallel} \citep{doParallel}, \texttt{dplyr} \citep{dplyr}, \texttt{ggplot2} \citep{ggplot2}. The reproducible R codes are available in the Supporting Information. An R package ``FSMT'' that implements the proposed method has been developed and is available at \url{https://github.com/YH-Mou/FSMT}.

\begin{table}[htbp]
  \centering
  \caption{
  Settings of treatment effect sizes and correlations between survival time and time-to-hospitalization endpoints in eight simulation scenarios}
    \begin{tabular}{lllr}
    \toprule
    \multicolumn{1}{c}{\multirow{2}[2]{*}{Scenario}} & \multicolumn{2}{c}{Treatment Effect} & \multicolumn{1}{c}{\multirow{2}[2]{*}{Kendall's Concordance}} \\
          & Survival Time & Time to Hospitalization &  \\
    \midrule
    S0    & None  & None  & 0 or 0.5 \\
    S1    & None  & Modest & 0.5 \\
    S2    & None  & Modest & 0 \\
    S3    & Modest & None  & 0.5 \\
    S4    & Modest & None  & 0 \\
    S5    & Very weak & Weak  & 0.5 \\
    S6    & Very weak & Weak  & 0 \\
    S7    & Weak  & Very Weak & 0.5 \\
    S8    & Weak  & Very Weak & 0 \\
    \bottomrule
    \end{tabular}%
  \label{tab:sim sce}%
\end{table}%

\subsection{Performance of FS-AT} 

\begin{figure*}
\centerline{\includegraphics[width=6in]{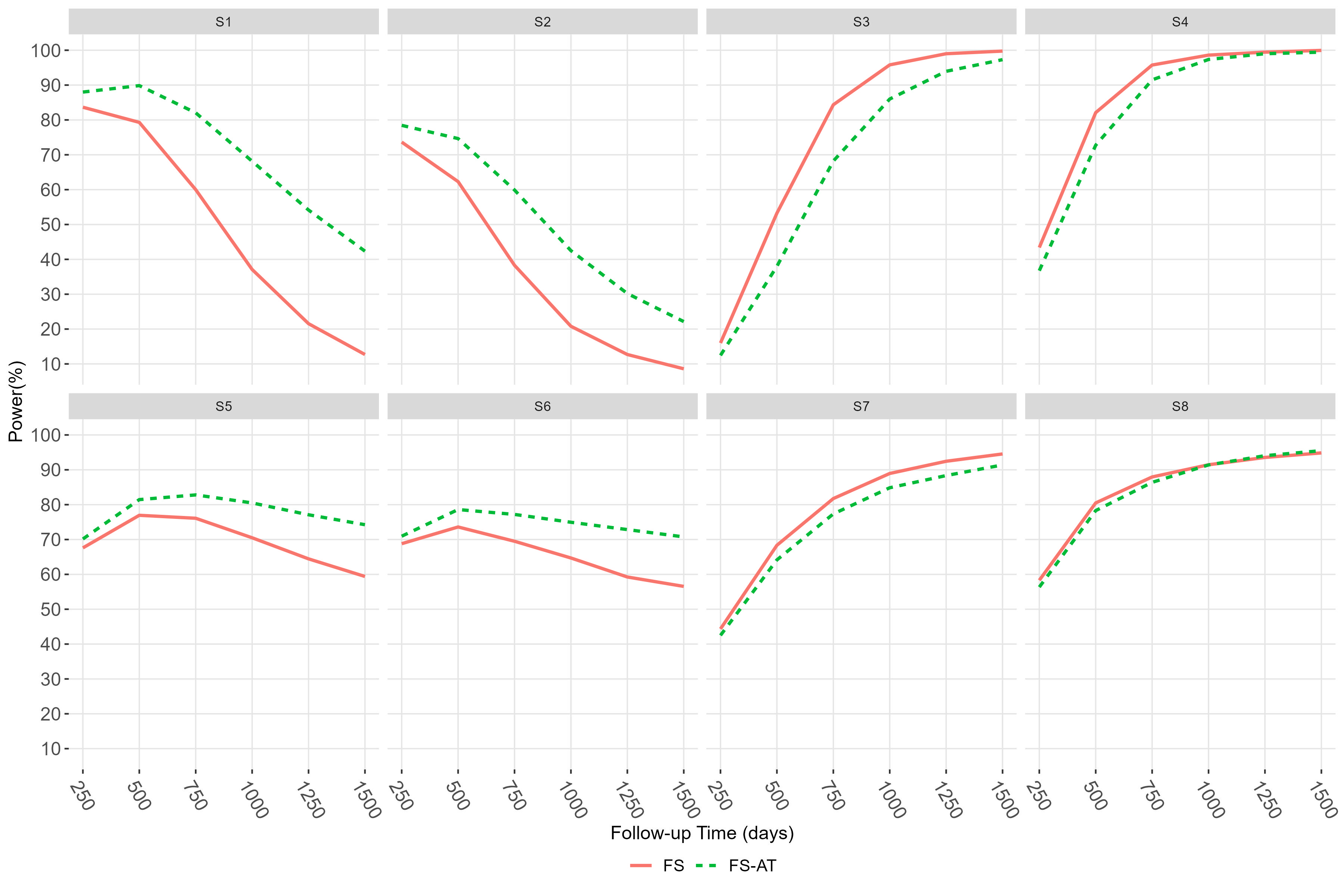}}
\caption{Empirical power of FS-AT and FS tests in simulation scenario: (S1) modest treatment effect on the time-to-hospitalization endpoint only, assuming a correlation between survival time and time-to-hospitalization endpoints, (S2) modest treatment effect on the time-to-hospitalization endpoint only, assuming no correlation, (S3) modest treatment effect on the survival time endpoint only, assuming a correlation, (S4) modest treatment effect on the survival time endpoint only, assuming no correlation, (S5) very weak treatment effect on the survival time endpoint and weak treatment effect on the time-to-hospitalization endpoint, assuming a correlation, (S6) very weak treatment effect on the survival time endpoint and weak treatment effect on the time-to-hospitalization endpoint, assuming no correlation, (S7) weak treatment effect on the survival time endpoint and very weak treatment effect on the time-to-hospitalization endpoint, assuming correlation, (S8) weak treatment effect on the survival time endpoint and very weak treatment effect on the time-to-hospitalization endpoint, assuming no correlation. \label{fig: FS-MT power}}
\end{figure*}

The empirical power of FS and FS-AT tests under simulation scenarios S1-S8 (i.e., under the non-null scenarios) is summarized in Figure \ref{fig: FS-MT power}. To begin with, in simulation scenarios S1 and S2, the FS-AT test continuously increases the power from the standard FS test by allowing the information on the time-to-hospitalization layer to contribute to the inference about the treatment effect. The difference between FS-AT and FS tests is more apparent with a longer follow-up time. Since longer follow-up time indicates more observed death events on the survival time layer, which brings more random differences, the FS-AT test increases the power by more frequently leveraging the time-to-hospitalization layer and bypassing the random difference on the survival time layer. With a shorter follow-up time, the FS test maintains good performance as lower death rates allow a higher contribution from the time-to-hospitalization layer to pairwise comparison results. In simulation scenarios S3 and S4, the treatment effect is on the survival time layer with random differences on the time-to-hospitalization layer. Although more frequently exploiting the time-to-hospitalization layer is expected to reduce power, the FS-AT test presents comparable power to the FS test in simulation scenario S4, due to employing the proposed adaptive thresholds. In simulation scenario S3, however, it is noticeable that the added correlation between endpoints lowers the power of both methods and the FS-AT test appears to be less powerful than the FS test. This phenomenon will be further investigated in Section \ref{se:inf of cor}.

In simulation scenarios S1-S4, we include only random differences on one of the endpoints to demonstrate the performance of FS-AT test under extreme situations. In practice, however, if prior information indicates no treatment effect on an endpoint, including it in the hierarchical structure should be discouraged when using either FS-AT or standard FS tests. In simulation scenarios S5-S8, we focus on the more common settings where there are treatment effects on both endpoints but with different magnitudes. In simulation scenarios S5 and S6, where the treatment effect on the survival time endpoint is lower than that on the time-to-hospitalization endpoint, the FS-AT test becomes more powerful than the FS test. Similar to simulation scenarios S1 and S2, a longer follow-up time may reduce power as more pairwise comparison results are determined by the survival time endpoint, for which the treatment effect has a weak signal. The FS-AT test appears to be more robust in terms of maintaining power in such a scenario. Different from the trends observed in simulation scenarios S1 and S2, the power of both FS and FS-AT tests here first increases from 250-day follow-up to 500-day follow-up and then starts to decrease, which is likely a result of the difference between having very weak or no treatment effects on the survival time layer. Additionally, in simulation scenarios S7 and S8 where the treatment effect on the time-to-hospitalization endpoint is weak, the FS-AT test achieves comparable power with the FS test (similar to simulation scenario S4). Finally, the empirical type I error in simulation scenario S0 with Kendall’s concordance set to 0 and 0.5 is presented in Table \ref{tab:sim s0}. At the nominal level of $\alpha = 0.05$, both FS and FS-AT tests adequately control empirical type I error rates, with observed values falling within the expected range of Monte Carlo variation.

\begin{table}[htbp!]
  \centering
  \caption{
 Empirical type I error rates (\%) obtained by FS-AT and FS under the null (simulation scenario S0) with Kendall’s concordance 0 and 0.5. The acceptable range of empirical type I error rate with Monte Carlo variation under 5000 replicates is 4.41\% to 5.64\% for the 5\% nominal level.}
    \begin{tabular}{lrrrr}
    \toprule
    Scenario & \multicolumn{1}{l}{Kendall’s Concordance} & \multicolumn{1}{l}{$\text{FU}$} & \multicolumn{1}{l}{FS} & \multicolumn{1}{l}{FS-AT} \\
    \midrule
    S0    & 0     & 500   & 5.12  & 5.16 \\
    S0    & 0     & 1000  & 5.24  & 5.34 \\
    S0    & 0     & 1500  & 4.96  & 5.10 \\
    S0    & 0.5   & 500   & 4.76  & 4.74 \\
    S0    & 0.5   & 1000  & 4.72  & 4.72 \\
    S0    & 0.5   & 1500  & 4.66  & 4.70 \\
    \bottomrule
    \end{tabular}%
  \label{tab:sim s0}%
\end{table}%

\subsection{Influence of the correlation} \label{se:inf of cor}

In this subsection, we present additional insights into the observed difference in power between simulation scenarios S3 and S4 
by examining how each endpoint or stage contributes through decomposition matrices.
The decomposition procedure presents the wins, losses, and ties for the treatment group participants in each stage.
In such a decomposition matrix, the percentages are all calculated based on the total number of pairwise comparisons between two contrasting participants. Then the calculation of three GPC measures at the stage level, NB, WR, and WO, follows.
Results with 750-day follow-up time averaged across 2000 replicates are shown in Table \ref{tab:decompose fs}. Results from other follow-up times are included in Supporting Information S3. 

In our example, the correlation between endpoints is more likely to influence the overall testing through the time-to-hospitalization endpoint, since the survival time endpoint is prioritized and therefore less likely to be affected by the correlation at the first stage.
To evaluate the contribution of the time-to-hospitalization endpoint in the FS test, we naturally consider the only stage $\mathscr{T}_{ij1}(0)$, which directly reflects this component’s effect. For the FS-AT test, between the stages $\mathscr{T}_{ij1}(t_1)$ and $\mathscr{T}_{ij1}(0)$, we give more attention to $\mathscr{T}_{ij1}(t_1)$ stage, where more ties are resolved by the time-to-hospitalization endpoint, leading to a greater influence on the overall testing outcome.
In simulation scenario S3, NB is negative and both WO and WR are less than one on these stages, suggesting potential negative stage-level treatment effects, whereas in simulation scenario S4, they are very close to zero or one, indicating no treatment effects. Importantly, in the FS test, the difference between approximately -6.10\% in simulation scenario S3 and -0.10\% in simulation scenario S4 for stage-level NB, 0.87 and 1.00 for stage-level WR, 0.89 and 1.00 for stage-level WO when averaged across 2000 replicates, is non-negligible. Similar magnitudes of difference are also observed in the FS-AT test, indicating that both the FS and FS-AT tests are influenced by the correlation structure. We next discuss the underlying mechanism driving the observed decrease in power in simulation scenario S3.

In simulation scenario S3, the marginal correlation between the survival time and time-to-hospitalization endpoints influences the chances to actually compare the time-to-hospitalization endpoint (conditioning on receiving ties in the survival time layer). As a result, although there is no treatment effect on the time-to-hospitalization endpoint marginally, the spurious, negative ``treatment effects'' are observed based on the pairwise comparisons between participants with no difference on the survival time layer. Such spurious negative ``treatment effects'' are the core reason for the reduced power in detecting the true positive treatment effects with both FS and FS-AT tests. These results underscore the importance of carefully considering the inclusion of additional endpoints in a hierarchical testing structure, as inter-endpoint correlations can affect treatment effect detection in unanticipated ways \citep{verbeeck2019generalized}. An endpoint that appears neutral with respect to power may, after conditioning on higher-layer outcomes, exhibit spurious negative or positive ``treatment effects'', potentially affecting statistical inference.
Additionally, Table \ref{tab:decompose fs} shows that the final stages of FS and FS-AT tests leave the same proportion of ties (3.15\% in simulation scenario S3 and 1.34\% in S4, respectively). This confirms that employing the last two stages of the FS-AT test prevents introducing additional ties and ensures full utilization of information, as in the standard FS test.

\begin{table}[htbp]
  \centering
  \caption{
  Stage-level decomposition of wins, losses, and ties in simulation scenarios S3 and S4 with 750-day follow-up time averaged across 2000 replicates. $\mathscr{D} _{ij}(d)$ and $\mathscr{T} _{ij1}(t_1)$ represent comparison functions for survival time and time-to-hospitalization endpoints with thresholds $d$ and $t_1$, respectively. Similar for $\mathscr{D} _{ij}(0)$ and $\mathscr{T} _{ij1}(0)$. The stage-level NB/WO/WR with the bolded values correspond to those relevant to the time-to-hospitalization endpoint. 
  Both simulation scenarios S3\&S4 feature modest treatment effects on the survival time endpoint only, where simulation scenario S3 assumes a correlation between endpoints and simulation scenario S4 assumes no such correlation.
  }
    \begin{tabular}{cclrrrrrr}
    \toprule
    \multirow{2}[2]{*}{Scenario} & \multirow{2}[2]{*}{Method} & \multicolumn{1}{c}{\multirow{2}[2]{*}{Stage}} & \multicolumn{1}{c}{\multirow{2}[2]{*}{Win(\%)}} & \multicolumn{1}{c}{\multirow{2}[2]{*}{Tie(\%)}} & \multicolumn{1}{c}{\multirow{2}[2]{*}{Loss(\%)}} & \multicolumn{3}{c}{Stage-level} \\
          &       &       &       &       &       & \multicolumn{1}{l}{NB(\%)} & \multicolumn{1}{l}{WO} & \multicolumn{1}{l}{WR} \\
    \midrule
    \multirow{6}[4]{*}{S3} & \multirow{2}[2]{*}{FS} & $\mathscr{D} _{ij}(0)$ & 37.29 & 35.15 & 27.56 & 9.72  & 1.22  & 1.35 \\
          &       & $\mathscr{T} _{ij1}(0)$ & 14.93 & 3.15  & 17.07 & \textbf{-6.10} & \textbf{0.89} & \textbf{0.87} \\
\cmidrule{2-9}          & \multirow{4}[2]{*}{FS-AT} & $\mathscr{D} _{ij}(d)$ & 30.27 & 48.01 & 21.71 & 8.56  & 1.19  & 1.39 \\
          &       & $\mathscr{T} _{ij1}(t_1)$ & 16.00 & 13.28 & 18.73 & \textbf{-5.69} & \textbf{0.89} & \textbf{0.85} \\
          &       & $\mathscr{D} _{ij}(0)$ & 2.48  & 8.76  & 2.04  & 3.29  & 1.07  & 1.21 \\
          &       & $\mathscr{T} _{ij1}(0)$ & 2.77  & 3.15  & 2.85  & \textbf{-0.90} & \textbf{0.98} & \textbf{0.97} \\
    \midrule
    \multirow{6}[4]{*}{S4} & \multirow{2}[2]{*}{FS} & $\mathscr{D} _{ij}(0)$ & 37.22 & 35.18 & 27.60 & 9.62  & 1.21  & 1.35 \\
          &       & $\mathscr{T} _{ij1}(0)$ & 16.90 & 1.34  & 16.94 & \textbf{-0.10} & \textbf{1.00} & \textbf{1.00} \\
\cmidrule{2-9}          & \multirow{4}[2]{*}{FS-AT} & $\mathscr{D} _{ij}(d)$ & 30.20 & 48.04 & 21.76 & 8.43  & 1.18  & 1.39 \\
          &       & $\mathscr{T} _{ij1}(t_1)$ & 18.31 & 11.39 & 18.34 & \textbf{-0.08} & \textbf{1.00} & \textbf{1.00} \\
          &       & $\mathscr{D} _{ij}(0)$ & 2.39  & 6.86  & 2.14  & 2.23  & 1.05  & 1.12 \\
          &       & $\mathscr{T} _{ij1}(0)$ & 2.76  & 1.34  & 2.76  & \textbf{-0.02} & \textbf{1.00} & \textbf{1.00} \\
    \bottomrule
    \end{tabular}%
  \label{tab:decompose fs}%
\end{table}%

\subsection{Influence of the caliper} \label{sec: inf of caliper}

In this subsection, we demonstrate the influence of caliper choices on adaptive thresholds. In addition to the default caliper ($c=20\%$), we consider $c=10\%, 40\%$, and the combined calipers. The combined caliper is an eight-stage FS-MT test that progressively employs $c=40\%,20\%,10\%$ calipers:
$\text{FS-MT} \left(
q_D^{40\%},q_1^{40\%}, q_D^{20\%},q_1^{20\%},
q_D^{10\%},q_1^{10\%}, 0,0
\right).$
The results are presented in Table \ref{tab:caliper}. In this simulation, a larger caliper is more effective at bypassing random differences, while a smaller caliper is more sensitive to detecting treatment effects on the survival time layer. The combined caliper is slightly enhanced by the additional stages used to bypass random differences. When increasing the caliper from 10\% to 20\% and 40\%, the performance gains in simulation scenarios S1, S2, S5, and S6, where treatment effects primarily occur on the time-to-hospitalization layer, generally outweigh the performance losses observed in the remaining scenarios where effects are concentrated on the survival time layer. This is expected since the adaptive thresholds work to keep detecting true signals while bypassing random differences. The combined caliper approach achieves the highest power under simulation scenarios S1, S2, S5, and S6, which is likely a result of its enhanced comparison structure that is more effective at bypassing the random differences. In general, the FS-AT test remains robust to the choice of caliper values as no rapid change in the performance of FS-AT test with different calipers is observed under most simulation scenarios.

\begin{table}[htbp]
  \centering
  \caption{
  Empirical power (\%) obtained by FS-AT with different calipers $c$ in simulation scenarios S1-S8. Notation: $\text{FU}$ represents the schedule follow-up length; Caliper-10\% refers to the FS-AT test with caliper $c=10\%$, similar for 20\% and 40\%; Combined Caliper refers to an eight-stage FS-AT test that progressively employs $c=40\%,20\%,10\%$ calipers.}
    \begin{tabular}{lrrrrr}
    \toprule
    Scenario & \multicolumn{1}{l}{$\text{FU}$} & \multicolumn{1}{l}{Caliper-10\%} & \multicolumn{1}{l}{Caliper-20\%} & \multicolumn{1}{l}{Caliper-40\%} & \multicolumn{1}{l}{Combined Caliper} \\
    \midrule
    S1    & 500   & 86.75 & 89.85 & 89.65 & 92.95 \\
    S1    & 1000  & 56.10 & 68.15 & 72.55 & 82.70 \\
    S1    & 1500  & 28.75 & 42.35 & 52.25 & 66.90 \\
    S2    & 500   & 70.35 & 74.65 & 80.55 & 83.00 \\
    S2    & 1000  & 31.35 & 42.50 & 60.90 & 65.40 \\
    S2    & 1500  & 14.90 & 22.15 & 39.85 & 45.15 \\
    S3    & 500   & 42.95 & 38.00 & 38.85 & 31.45 \\
    S3    & 1000  & 90.85 & 86.00 & 83.25 & 74.25 \\
    S3    & 1500  & 98.90 & 97.30 & 95.30 & 91.60 \\
    S4    & 500   & 77.00 & 72.75 & 69.25 & 66.30 \\
    S4    & 1000  & 98.10 & 97.35 & 95.45 & 94.35 \\
    S4    & 1500  & 99.70 & 99.45 & 98.95 & 98.95 \\
    S5    & 500   & 80.10 & 81.45 & 81.40 & 83.35 \\
    S5    & 1000  & 76.80 & 80.50 & 82.00 & 84.45 \\
    S5    & 1500  & 68.90 & 74.25 & 77.75 & 80.80 \\
    S6    & 500   & 76.40 & 78.60 & 80.80 & 81.70 \\
    S6    & 1000  & 70.20 & 74.95 & 80.65 & 82.15 \\
    S6    & 1500  & 63.20 & 70.75 & 77.70 & 80.10 \\
    S7    & 500   & 65.60 & 64.15 & 65.10 & 62.40 \\
    S7    & 1000  & 86.50 & 84.85 & 84.25 & 81.25 \\
    S7    & 1500  & 92.85 & 91.40 & 90.25 & 88.10 \\
    S8    & 500   & 78.95 & 78.30 & 77.05 & 76.05 \\
    S8    & 1000  & 91.65 & 91.40 & 90.00 & 89.70 \\
    S8    & 1500  & 95.65 & 95.50 & 95.65 & 95.65 \\
    \bottomrule
    \end{tabular}%
  \label{tab:caliper}%
\end{table}%

\subsection{Influence of the weight}

In this subsection, we show the influence of weight choices on adaptive thresholds. Alternative to the default weight choice $w=1$, we also examine $w\in\{0.1,0.3,0.5\}$. In the resulting four candidate tests, FS-AT(1), FS-AT(0.5), FS-AT(0.3), FS-AT(0.1), a smaller $w$ places greater emphasis on the survival time endpoint. According to the results presented in Table \ref{tab:weight}, in simulation scenarios S1, S2, S5, and S6, where the treatment effects are primarily on the time-to-hospitalization layer, smaller $w$ leads to lower power. This is not surprising since emphasizing the higher-layer endpoint naturally reduces the chance in detecting the treatment effect on the lower layer. In simulation scenarios S3, S4, S7, and S8, where the treatment effects are primarily on the survival time layer, relatively more comparable power is achieved by different choices of $w$. These indicate that the FS-AT test is generally robust to the choice of $w$ in terms of maintaining the capacity to detect treatment effects on the survival time layer.

\begin{table}[htbp]
  \centering
  \caption{Empirical power (\%) obtained by FS-AT tests with different weights $w$ in simulation scenarios S1-S8. Notation: $\text{FU}$ represents the schedule follow-up length; FS-AT(0.1) refers to the FS-AT test with weight $w=0.1$, similar for FS-AT(0.3), FS-AT(0.5), and FS-AT(1.0).}
    \begin{tabular}{lrrrrr}
    \toprule
    Scenario & \multicolumn{1}{l}{$\text{FU}$} & \multicolumn{1}{l}{FS-AT(0.1)} & \multicolumn{1}{l}{FS-AT(0.3)} & \multicolumn{1}{l}{FS-AT(0.5)} & \multicolumn{1}{l}{FS-AT(1.0)} \\
    \midrule
    S1    & 500   & 79.30 & 82.50 & 87.00 & 89.85 \\
    S1    & 1000  & 37.10 & 51.10 & 59.50 & 68.15 \\
    S1    & 1500  & 13.10 & 24.70 & 32.55 & 42.35 \\
    S2    & 500   & 62.30 & 71.05 & 73.35 & 74.65 \\
    S2    & 1000  & 22.45 & 36.10 & 39.80 & 42.50 \\
    S2    & 1500  & 10.75 & 16.85 & 19.55 & 22.15 \\
    S3    & 500   & 53.25 & 49.20 & 43.50 & 38.00 \\
    S3    & 1000  & 95.80 & 92.80 & 90.25 & 86.00 \\
    S3    & 1500  & 99.75 & 99.20 & 98.50 & 97.30 \\
    S4    & 500   & 82.10 & 77.25 & 74.45 & 72.75 \\
    S4    & 1000  & 98.55 & 97.95 & 97.55 & 97.35 \\
    S4    & 1500  & 99.80 & 99.60 & 99.45 & 99.45 \\
    S5    & 500   & 76.95 & 79.10 & 80.60 & 81.45 \\
    S5    & 1000  & 70.50 & 75.30 & 77.95 & 80.50 \\
    S5    & 1500  & 60.25 & 67.50 & 71.65 & 74.25 \\
    S6    & 500   & 73.60 & 76.95 & 78.10 & 78.60 \\
    S6    & 1000  & 65.35 & 72.35 & 74.10 & 74.95 \\
    S6    & 1500  & 59.70 & 67.25 & 69.10 & 70.75 \\
    S7    & 500   & 68.35 & 67.40 & 65.75 & 64.15 \\
    S7    & 1000  & 88.95 & 87.70 & 86.40 & 84.85 \\
    S7    & 1500  & 94.35 & 93.35 & 92.60 & 91.40 \\
    S8    & 500   & 80.50 & 79.45 & 78.50 & 78.30 \\
    S8    & 1000  & 91.30 & 91.60 & 91.60 & 91.40 \\
    S8    & 1500  & 95.15 & 95.50 & 95.50 & 95.50 \\
    \bottomrule
    \end{tabular}%
  \label{tab:weight}%
\end{table}%

\section{Case Study: Digitalis Investigation Group (DIG) clinical trial}\label{data_example}
To illustrate the proposed testing procedure, we apply the FS-AT test to analyze the Digitalis Investigation Group (DIG) trial (ClinicalTrials.gov identifier: NCT00000476). This clinical trial investigated the effect of digoxin on patients with heart failure. Patients who had heart failure and a left ventricular ejection fraction of 0.45 or less were eligible for the primary randomization. This study was neutral for its primary outcome of all-cause mortality but evaluated both death and hospitalization events. When these two outcomes were analyzed separately, treatment effects appeared mainly on the hospitalization outcomes. Alternatively, we hereby employ FS-AT and standard FS tests to reanalyze this dataset. In this case study, both death and hospitalization outcomes are analyzed in the time-to-event form with the right censoring information taken into account (i.e., survival time and time to hospitalization). In developing the endpoint hierarchy, time to (all-cause) death is included as the higher layer, and time to (all-cause) hospitalization is included as the lower layer in both FS-AT and FS tests. For illustration of the differential operating characteristics between these tests, we focus on key subgroups from the trial population and use stratification for better adjustment in this analysis. That is, we study the subgroup population with New York Heart Association (NYHA) functional class III or IV and stratify by ejection fraction ($<0.25$ or $0.25-0.45$), cause of heart failure (ischemic or nonischemic), and age ($<70$ or $\geq 70$). The number of patients within each stratum is presented in Supporting Information S4. For the FS-AT test, the adaptive thresholds with default weight and caliper ($w=1,c=20\%$) are adopted. 
In both approaches, the stratification step follows the details presented in \citet{finkelstein1999combining}. 
Specifically, we first divide patients into distinct strata indexed by $r = 1, 2, \dots, R$ ($R=8$ in our case study with the DIG trial), and let $\mathcal{A}_r$ denote the set of indices for the $n_r$ participants in stratum $r$. For each individual $i \in \mathcal{A}_r$, the individual score is computed within stratum as  
$
U_i = \sum_{j \in \mathcal{A}_r} U_{ij}.
$
The stratified test statistic then takes the form  
$
S_{\text{strat}} = \sum_{r=1}^{R} \sum_{i \in \mathcal{A}_r} Z_i U_i,
$
where $Z_i$ indicates treatment assignment. The variance of $S_{\text{strat}}$ is estimated analogously to the unstratified case by aggregating within-stratum variances:  
$
\widehat{\text{Var}}(S_{\text{strat}}) = \sum_{r=1}^{R} 
\frac{m_r(n_r - m_r)}{n_r(n_r - 1)} 
\left( \sum_{i \in \mathcal{A}_r} U_i^2 \right),
$
where $m_r$ is the number of treated participants in stratum $r$. A summary of alternative stratification approaches is provided in \citet{backer2025stratification}.

The resulting p-values are 0.033 and 0.055 for FS-AT and FS tests, respectively. Compared with the FS test, the FS-AT test yields a smaller p-value and concludes a significant treatment effect at the 0.05 level. 
To understand the underlying driver of such a difference, we present decomposition matrices and stage-level contribution to the overall hypothesis testing, quantified by NB, WO, and WR in Table~\ref{tab:DIG_decompose}.
In this stratified analysis, the decomposition is obtained in a manner consistent with the stratified testing procedure. Specifically, within each stratum, we compute the counts of wins, ties, and losses from pairwise comparisons between treatment and control participants. These counts are then summed across all strata to yield the overall numbers of wins ($N_+$), ties ($N_0$), and losses ($N_-$). The corresponding proportions, expressed as percentages of the total number of pairwise comparisons between treatment and control groups, form the stage-level decomposition results presented in the table. These values are then used to calculate the stage-level GPC measures, which are reported to aid interpretation of each endpoint’s contribution to the overall test.
It is noticeable that, the treatment effect is weak on the survival time endpoint but relatively stronger on the time-to-hospitalization endpoint.
For instance, in the FS test, the stage-level NB is approximately -0.77\% for the survival time endpoint and 5.09\% for the time-to-hospitalization endpoint, indicating a noticeably stronger treatment effect in favor of the treatment group at the latter stage. A similar pattern is observed in the FS-AT test, where the corresponding NB values are -0.48\% and 4.92\% at the first two stages, when most ties were resolved. Consistent trends are also reflected in WO and WR.
Therefore, the standard FS test may be less powerful as it assigns the strict priority given to the survival time. In contrast, the FS-AT test demonstrates a higher capacity in detecting the treatment effect under such cases. Finally, we report p-values for the hypothesis tests and interpret statistical significance at the 0.05 level only for the purposes of demonstration. Nevertheless, we recognize the caveat that, in practice, study conclusions may not depend solely on whether a p-value is greater than or less than a single cutoff point like 0.05.

\begin{table}[htbp]
  \centering
  \caption{
  Stage-level decomposition of wins, losses, and ties in the DIG trial case study. $\mathscr{D} _{ij}(d)$ and $\mathscr{T} _{ij1}(t_1)$ represent comparison functions for time to all-cause death and time to all-cause hospitalization endpoints with thresholds $d$ and $t_1$, respectively. Similar for $\mathscr{D} _{ij}(0)$ and $\mathscr{T} _{ij1}(0)$. The FS test has a p-value of 0.055 and the FS-AT test has a p-value of 0.033.
  }
    \begin{tabular}{clrrrrrr}
    \toprule
    \multirow{2}[2]{*}{Method} & \multicolumn{1}{c}{\multirow{2}[2]{*}{Stage}} & \multicolumn{1}{c}{\multirow{2}[2]{*}{Win(\%)}} & \multicolumn{1}{c}{\multirow{2}[2]{*}{Tie(\%)}} & \multicolumn{1}{c}{\multirow{2}[2]{*}{Loss(\%)}} & \multicolumn{3}{c}{Stage-level} \\
          &       &       &       &       & \multicolumn{1}{l}{NB(\%)} & \multicolumn{1}{l}{WO} & \multicolumn{1}{l}{WR} \\
    \midrule
    \multirow{2}[2]{*}{FS} & $\mathscr{D} _{ij}(0)$ & 17.33 & 64.58 & 18.10 & -0.77 & 0.98  & 0.96 \\
          & $\mathscr{T} _{ij1}(0)$ & 29.60 & 10.47 & 24.51 & 5.09  & 1.17  & 1.21 \\
    \midrule
    \multirow{4}[2]{*}{FS-AT} & $\mathscr{D} _{ij}(d)$ & 11.20 & 77.13 & 11.67 & -0.48 & 0.99  & 0.96 \\
          & $\mathscr{T} _{ij1}(t_1)$ & 28.21 & 25.64 & 23.28 & 4.92  & 1.14  & 1.21 \\
          & $\mathscr{D} _{ij}(0)$ & 1.34  & 22.94 & 1.36  & -0.02 & 1.00  & 0.99 \\
          & $\mathscr{T} _{ij1}(0)$ & 6.53  & 10.47 & 5.94  & 0.59  & 1.05  & 1.10 \\
    \bottomrule
    \end{tabular}%
  \label{tab:DIG_decompose}%
\end{table}%

\section{Discussion}\label{discussion}
In this study, we develop a hypothesis testing approach for composite time-to-event endpoints with a hierarchical structure. Modifying the standard FS test, the proposed FS-MT test adds layers with additional thresholds to pairwise comparisons. 
Weighted adaptive thresholds are also developed to reduce the potential challenge in determining the additional thresholds. With simulated datasets, we compare the performance of FS-AT and FS tests while addressing the influence of the follow-up time, treatment effect sizes, and correlations between the two endpoints. The numerical results show that the FS-AT test outperforms the FS test when the effect is primarily on the lower layer and generally maintains comparable power when the effect is primarily on the higher layer. 

In the FS-AT test, there are two tuning parameters: weight $w$ and caliper $c$. As a general rule of thumb, we recommend using $w=1,c=20\%$ as the default when clinical priors are not sufficiently strong. 
We have investigated several different choices of $c$ in the simulations to provide additional insights into their working mechanisms. 
In addition to different choices of $c$, the combined caliper approach is a strong alternative (e.g., $\text{FS-MT} \left(
q_D^{40\%},q_1^{40\%}, q_D^{20\%},q_1^{20\%},
q_D^{10\%},q_1^{10\%}, 0,0
\right)$ as employed in Section \ref{sec: inf of caliper}). The performance of the combined caliper approach tends to rely mainly on the largest caliper because the priority is given to this caliper by the specification of the hierarchy. It is also worth noting that extra attentions are given to the lower layer (i.e., time-to-hospitalization endpoint in our example) in this approach since each one of the added calipers will provide additional opportunities for the lower layer to contribute information to the pairwise comparison. In this combined caliper approach, the additional computation time brought by the added stages may need to be considered, particularly when the study sample size is large or repeated calculations are required (e.g., resampling-based inference). On the other hand, when $c$ is fixed, the weight $w\in [0,1]$ increases the adaptive threshold of the lower layer to address the higher layer endpoint. Since the reciprocal function in the $ q_1^{c}/w$ structure is not on a linear scale, care is needed to use a relatively small $w$. Apart from adopting the reciprocal function, one may also consider the two-caliper structure ($c_1,c_2$). This architecture employs a larger quantile to increase the lower layer threshold. When a small $w$ is desired, one may employ this approach and adjust $c_2$ to avoid the potential trivial stage when $w$ is specified close to zero. Despite the flexibility of the proposed adaptive threshold system, the structure ($w,c$) or ($c_1,c_2$) should be pre-specified. Finally, when domain knowledge allows for precisely specifying all thresholds, directly implementing the FS-MT test could potentially enhance the clinical interpretability of each stage. In general, the FS-AT test serves to simplify the process for selecting tuning parameters based on general rule of thumbs, when clinical information is insufficient to determine the thresholds.

We note that, whether using the FS or FS-MT (FS-AT) test, clinical relevance remains important in selecting and prioritizing endpoints in the outcome specification stage. In general, the FS-MT test still assumes that composite endpoints and their hierarchy are defined a priori based on clinical judgment. That is, the FS-MT test does not completely overturn this ordering; rather, it introduces a more flexible comparison mechanism that starts from the highest-priority endpoint but then permits evaluation of lower-tier endpoints only when differences on the top layer are within a non-informative range. Hence, the FS-MT test occupies a middle ground between strict hierarchical enforcement (the standard FS test) and complete re-ranking of endpoints. In practice, if strong clinical priors exist for fixed thresholds, they should be incorporated within the FS-MT test. In the absence of such priors, the FS-AT test offers a pragmatic alternative to adaptively define these thresholds. We have not argued the use of FS-MT and FS-AT tests to replace the standard FS test in all settings. In fact, when clinical relevance demands strict hierarchical testing, the standard FS test will still be preferred as primary analysis, but FS-MT and FS-AT tests may contribute useful information in secondary analysis. In any case, large discrepancies between standard FS and FS-MT (FS-AT) testing results should prompt reflection on what aspect of treatment benefit is most meaningful given the context of the specific clinical trial.

Although our focus has been on hypothesis testing with the FS test, the proposed comparison structure can also be used to estimate treatment effect measures within the GPC framework. By incorporating the developed win functions into the computation of $N_+$, $N_-$, and $N_0$, one can obtain corresponding treatment effect estimates, as shown in equation~(\ref{GPC_measures}). This formulation naturally extends beyond the NB, WO, and WR measures and can be adapted to other measures and methods within the GPC framework. While our study focuses on the FS test, we acknowledge that alternative hypothesis testing procedures within the GPC framework possess distinct operating characteristics and may outperform each other under certain conditions. For instance, to better handle censoring in time-to-event endpoints, the proposed win function could be incorporated into alternative approaches such as inverse probability of censoring weighting \citep{dong2020inverse, dong2021adjusting} or other methods specifically developed for censored data \citep{peron2018extension, deltuvaite2023generalized, deltuvaite-thomas2025rightcensored}. Addressing censoring in our setting with multiple thresholds is a valuable topic for future research.

\section*{Acknowledgement}
The authors thank the DIG study team and the DIG study team for making the data available through the Biologic Specimen and Data Repository Information Coordinating Center (BioLINCC) at the National Heart, Lung, and Blood Institute (NHLBI). F. Li is supported by the United States National Institutes of Health and the National Heart, Lung, and Blood Institute (grant number R01-HL168202). Funders had no role in study design, data collection, analysis, reporting or the decision to submit for publication.

\bibliographystyle{biom} 
\bibliography{refs}%

\newpage

{\LARGE \textbf{Supporting Information} }
\appendix

\renewcommand{\thesection}{S\arabic{section}}
\renewcommand{\thesubsection}{S\arabic{section}.\arabic{subsection}}
\renewcommand{\thetable}{S\arabic{table}}
\setcounter{table}{0}

\section{
Generalization of the FS-MT test with additional thresholds and multiple non-fatal events
}
In the main text, we introduce a 4-stage FS-MT test with survival time and time-to-hospitalization endpoints. For the ease of generalization, we first rewrite each stage in $\text{FS-MT}(d,t_1,0,0)$ with comparison functions:
\begin{enumerate}
\item[] \hspace{-0.15in} Stage 1: Compare the survival difference at the $d$ level. That is:
    \begin{equation*}
        U_{ij1} = \mathscr{D}_{ij}(d).
    \end{equation*}

\item[] \hspace{-0.15in} Stage 2: If $U_{ij1}=0$, compare time-to-hospitalization difference at the $t_1$ level. That is:
    \begin{equation*}
        U_{ij2} = I(U_{ij1}=0)\times \mathscr{T}_{ij1}(t_1) 
            = \left(1-\mathscr{D}_{ij}(d)\right)\left(1+\mathscr{D}_{ij}(d)\right)
                \mathscr{T}_{ij1}(t_1) 
            = \left(1-\mathscr{D}^2_{ij}(d)\right) \mathscr{T}_{ij1}(t_1).
    \end{equation*}

\item[] \hspace{-0.15in} Stage 3: If $U_{ij2}=0$, compare any survival difference between $i$ and $j$. That is:
    \begin{equation*}
        U_{ij3} = I(U_{ij1}=U_{ij2}=0)\times \mathscr{D}_{ij}(0) 
            = \left(1-\mathscr{D}^2_{ij}(d)\right) \left(1-\mathscr{T}^2_{ij1}(t_1)\right)
                \mathscr{D}_{ij}(0).
    \end{equation*}

\item[] \hspace{-0.15in} Stage 4: If $U_{ij3}=0$, compare any time-to-hospitalization difference between $i$ and $j$. That is:
    \begin{equation*}
        U_{ij4} = I(U_{ij1}=U_{ij2}=U_{ij3}=0)\times \mathscr{T}_{ij1}(0)
            = \left(1-\mathscr{D}^2_{ij}(d)\right) \left(1-\mathscr{T}^2_{ij1}(t_1)\right)
                \left(1-\mathscr{D}^2_{ij}(0)\right) \mathscr{T}_{ij1}(0).
    \end{equation*}
    
\item[] \hspace{-0.15in} Summing the above four components, we obtain the win-loss score for the pair $i$ and $j$ as (including the set of thresholds in the argument for the win-loss score):
    \begin{eqnarray*}
        U_{ij}(d,t_1,0,0) 
        & = & U_{ij1} + U_{ij2} + U_{ij3} + U_{ij4} \\
        & = &\mathscr{D}_{ij}(d) 
            + \left(1-\mathscr{D}^2_{ij}(d)\right) \mathscr{T}_{ij1}(t_1)
            + \left(1-\mathscr{D}^2_{ij}(d)\right) \left(1-\mathscr{T}^2_{ij1}(t_1)\right)
                \mathscr{D}_{ij}(0) \\ 
            && + \left(1-\mathscr{D}^2_{ij}(d)\right) \left(1-\mathscr{T}^2_{ij1}(t_1)\right)
                \left(1-\mathscr{D}^2_{ij}(0)\right) \mathscr{T}_{ij1}(0) \\
        & = &\mathscr{D}_{ij}(d) + \left(1-\mathscr{D}^2_{ij}(d)\right) 
                \big\{  
                    \mathscr{T}_{ij1}(t_1) 
                    + \left(1-\mathscr{T}^2_{ij1}(t_1)\right) \mathscr{D}_{ij}(0)\\
                    &&+ \left(1-\mathscr{T}^2_{ij1}(t_1)\right) \left(1-\mathscr{D}^2_{ij}(0)\right) \mathscr{T}_{ij1}(0)
                \big\} \\
        & = & \mathscr{D}_{ij}(d) + \left(1-\mathscr{D}^2_{ij}(d)\right) 
                \big\{
                    \mathscr{T}_{ij1}(t_1) + \left(1-\mathscr{T}^2_{ij1}(t_1)\right)\\
                    &&\times \left[ 
                        \mathscr{D}_{ij}(0) + \left(1-\mathscr{D}^2_{ij}(0)\right) \mathscr{T}_{ij1}(0)
                    \right]
                \big\}.
    \end{eqnarray*}
\end{enumerate}

With such two endpoints, we may consider having $L\geq 1$ non-zero thresholds for each of them, namely $\left\{ d_1,d_2,...,d_L > 0 \right\}$ for survival time and $\left\{ t_{11},t_{12},...,t_{1L} > 0 \right\}$ for the nonfatal event. Given these two series of thresholds, the win-loss score for the pair comparison between $i$ and $j$ in this $2(L+1)$-stage FS-MT test, FS-MT$(d_1,t_{11},d_2,t_{12},...,d_L,t_{1L},0,0)$ is:
\begin{eqnarray*}
    && U_{ij}(d_1,t_{11},d_2,t_{12},...,d_L,t_{1L},0,0) \\
    & = & \mathscr{D}_{ij}(d_1) 
        + \left(1-\mathscr{D}^2_{ij}(d_1) \right) \mathscr{T}_{ij1}(t_{11})
        + \left(1-\mathscr{D}^2_{ij}(d_1) \right) 
            \left( 1- \mathscr{T}^2_{ij1}(t_{11})\right) \mathscr{D}_{ij}(d_2) \\
        &+& \left(1-\mathscr{D}^2_{ij}(d_1) \right) 
            \left( 1- \mathscr{T}^2_{ij1}(t_{11})\right)
            \left(1-\mathscr{D}^2_{ij}(d_2) \right) \mathscr{T}_{ij1}(t_{12}) + ... \\
        &+& \prod_{l=1}^{L} \left[ \left(1-\mathscr{D}^2_{ij}(d_l) \right)
                \left( 1- \mathscr{T}^2_{ij1}(t_{1l})\right) \right] \mathscr{D}_{ij}(0) \\
        &+& \prod_{l=1}^{L} \left[ \left(1-\mathscr{D}^2_{ij}(d_l) \right)
                \left( 1- \mathscr{T}^2_{ij1}(t_{1l})\right) \right]
                \left(1-\mathscr{D}^2_{ij}(0) \right) \mathscr{T}_{ij1}(0).      
\end{eqnarray*}

In addition, it is possible to encounter more than one nonfatal events in a study. For the $K\geq 1$ nonfatal events, we define $T_{ki},(k=1,2,...,K)$ as the observed time to the nonfatal events (e.g., hospitalization, stroke, ischemia) of $i$, and $\Delta_{ki}$ as the corresponding censoring indicator for the $k$th nonfatal event. With one non-zero threshold $t_{k1}$ for each nonfatal event, the comparison functions for nonfatal events are then:
\begin{eqnarray*}
    \mathscr{T} _{ijk}(t_{k1}) &=& I \left(|T_{ki}-T_{kj}|\geq t_{k1} \right) \\
                        && \times 
                        [ (1-\Delta_{ki})(1-\Delta_{kj})\text{Sign}(T_{ki}-T_{kj})
                                + \Delta_{ki}(1-\Delta_{kj})I(T_{ki} \geq T_{kj})\\
                                &&\quad - (1-\Delta_{ki})\Delta_{kj}I(T_{ki} \leq T_{kj}) ]
                ,~~~k=1,\ldots,K.
\end{eqnarray*}

The FS-MT test formed by the survival time (with one non-zero threshold $d$) and $K$ nonfatal events can be decomposed into the following $2(K+1)$ components:
\begin{eqnarray*}
    && U_{ij1} = \mathscr{D}_{ij}(d), \\
    && U_{ij2} = \left(1-\mathscr{D}^2_{ij}(d) \right) \mathscr{T} _{ij1}(t_{11}), \\
    && U_{ij3} = \left(1-\mathscr{D}^2_{ij}(d) \right) 
                    \left( 1- \mathscr{T}^2_{ij1}(t_{11})\right)
                        \mathscr{T} _{ij2}(t_{21}), \\
    && ..., \\
    && U_{ij,K+1} = \left(1-\mathscr{D}^2_{ij}(d) \right) \prod_{k=1}^{K-1}
                    \left( 1- \mathscr{T}^2_{ijk}(t_{k1})\right) 
                    \mathscr{T} _{ijK}(t_{K1}), \\
    && ..., \\
    && U_{ij,2(K+1)} = \left(1-\mathscr{D}^2_{ij}(d) \right) \prod_{k=1}^{K}
                        \left( 1- \mathscr{T}^2_{ijk}(t_{k1})\right) 
                        \left(1-\mathscr{D}^2_{ij}(0) \right)
                        \prod_{k=1}^{K-1} \left( 1- \mathscr{T}^2_{ijk}(0)\right)
                        \mathscr{T}_{ijK}(0).                       
\end{eqnarray*}

By summing the above components, the final win-loss score for the pair comparison between $i$ and $j$ is:
\begin{equation*}
    U_{ij}(d,t_{11},t_{21},...,t_{K1},0,...,0) = \sum_{s=1}^{2(K+1)} U_{ijs}.
\end{equation*}

Based on the above two scenarios, in the most general scenario, for survival time and $K\geq 1$ nonfatal events, we may specify $L\geq 1$ non-zero thresholds for each of them to form the $(K+1)(L+1)-$stage FS-MT test. Denote non-zero thresholds for survival time as $\left\{ d_1,d_2,...,d_L > 0 \right\}$, and for the $k$th nonfatal event as $\left\{ t_{k1},t_{k2},...,t_{kL} > 0 \right\}$. The general expression for the win-loss score of pair comparison between $i$ and $j$ becomes:
\begin{eqnarray*}
    && U_{ij}(d_1,t_{11},t_{21},...,t_{K1},
              d_2,t_{12},t_{22},...,t_{K2},...,
              d_L,t_{1L},t_{2L},...,t_{KL},0,...,0) \\
    &=& \mathscr{D}_{ij}(d) 
        + \left(1-\mathscr{D}^2_{ij}(d_1) \right) \mathscr{T} _{ij1}(t_{11}) + ... 
        + \left(1-\mathscr{D}^2_{ij}(d_1) \right) \prod_{k=1}^{K-1}
                    \left( 1- \mathscr{T}^2_{ijk}(t_{k1})\right) 
                    \mathscr{T} _{ijK}(t_{K1}) + ... \\
        &+& \left(1-\mathscr{D}^2_{ij}(d_1) \right)\left(1-\mathscr{D}^2_{ij}(d_2) \right)
                    \prod_{k=1}^{K} \left( 1- \mathscr{T}^2_{ijk}(t_{k1})\right) 
                    \prod_{k=1}^{K-1} \left( 1- \mathscr{T}^2_{ijk}(t_{k2})\right) \mathscr{T}_{ijk}(t_{K2}) + ... \\
        &+& \prod_{l=1}^{L} \left(1-\mathscr{D}^2_{ij}(d_l) \right) 
            \prod_{l=1}^{L} \prod_{k=1}^{K} \left( 1- \mathscr{T}^2_{ijk}(t_{kl})\right)
            \left(1-\mathscr{D}^2_{ij}(0) \right) \prod_{k=1}^{K-1} 
            \left( 1- \mathscr{T}^2_{ijk}(0)\right) \mathscr{T}_{ijK}(0).           
\end{eqnarray*}

The adaptive thresholds proposed in Section 2.3 are also applicable in more general settings. For the survival time and $K$ nonfatal events, we define $K+1$ sequences of all non-zero differences in pair comparisons as:
\begin{eqnarray*}
   \mathbb{D} &=& \left\{ |D_i - D_j| : \forall i,j \in \{1,2,...,n\}, 
                    (D_i - D_j) \neq 0   \right\}, \\
   \mathbb{T}_k &=& \left\{ |T_{ki} - T_{kj}| : \forall i,j \in \{1,2,...,n\}, 
                    (T_{ki} - T_{kj}) \neq 0   \right\},~~~k=1,\ldots,K.
\end{eqnarray*}
The empirical $c$ quantile values of these sequences are then $q_D^{c},q_1^{c},q_2^{c},...,q_K^{c}$. With $K$ weights for the nonfatal events $\{w_1,w_2,..,w_K\}$, we have FS-AT$(w_1,w_2,..,w_K)^{c}$ as FS-MT$(d=q_D^{c},t_{11}=q_1^{c}/w_1, t_{21}=q_2^{c}/w_2,..., t_{K1}=q_K^{c}/w_K)$.

\section{Additional simulation scenarios} \label{apx: add sim sce}
In this appendix, we include simulation results for 8 additional simulation scenarios S9-S16, as listed in Table \ref{tab:apx scenario}. The simulation setup is the same as that in the main text for simulation scenarios S1-S8. The table version of the power achieved by FS-AT and FS tests in simulation scenarios S1-S8 is also included.

\begin{table}[htbp]
  \centering
  \caption{Settings of treatment effect sizes and correlations between survival time and time-to-hospitalization endpoints in eight additional simulation scenarios}
    \begin{tabular}{lllr}
    \toprule
    \multicolumn{1}{c}{\multirow{2}[2]{*}{Scenario}} & \multicolumn{2}{c}{Treatment Effect} & \multicolumn{1}{c}{\multirow{2}[2]{*}{Kendall's Concordance}} \\
          & Survival Time & Time to Hospitalization &  \\
    \midrule
    S9    & Weak  & None  & 0.5 \\
    S10   & Weak  & None  & 0 \\
    S11   & Weak  & Weak  & 0.5 \\
    S12   & Weak  & Weak  & 0 \\
    S13   & Weak  & Modest & 0.5 \\
    S14   & Weak  & Modest & 0 \\
    S15   & Very weak & Very weak & 0.5 \\
    S16   & Very weak & Very weak & 0 \\
    \bottomrule
    \end{tabular}%
  \label{tab:apx scenario}%
\end{table}%

\subsection{Performance of FS-AT}

In simulation scenarios S9\&S10, where the weak treatment effect is on the survival time endpoint only, compared to the FS test, the FS-AT test achieves lower but comparable power when correlation is not assumed, and the difference in power is larger with correlation assumed. Such observation is similar to that in simulation scenarios S3\&S4. In simulation scenarios S11-S14, FS-AT and FS tests perform well as the treatment effects are on both endpoints and the overall strength of such effects is strong enough. In simulation scenarios S15\&S16, where very weak treatment effects are on both endpoints, the FS-AT test achieves comparable but slightly higher power than the FS test. This is likely a result of providing more chances for the time-to-hospitalization endpoint, where the treatment effect is easier to observe, as the administrative censoring rate is usually lower than that for the survival time endpoint.

\begin{longtable}[c]{lrrr|lrrr}
\captionsetup{justification=centering}
\caption{Empirical power (\%) obtained by FS-AT and FS tests in simulation scenarios S1-S16} \\
\toprule
\textbf{Scenario} & \multicolumn{1}{l}{FU} & \multicolumn{1}{l}{FS} & \multicolumn{1}{l|}{FS-AT} & \textbf{Scenario} & \multicolumn{1}{l}{FU} & \multicolumn{1}{l}{FS} & \multicolumn{1}{l}{FS-AT} \\
\midrule
\endfirsthead

\toprule
\textbf{Scenario} & \multicolumn{1}{l}{FU} & \multicolumn{1}{l}{FS} & \multicolumn{1}{l|}{FS-AT} & \textbf{Scenario} & \multicolumn{1}{l}{FU} & \multicolumn{1}{l}{FS} & \multicolumn{1}{l}{FS-AT} \\
\midrule
\endhead

\midrule
\multicolumn{8}{r}{\textit{Continued on next page}} \\
\midrule
\endfoot

\bottomrule
\endlastfoot

    S1    & 250   & 83.65 & 88.00 & S9    & 250   & 10.60 & 9.00 \\
    S1    & 500   & 79.30 & 89.85 & S9    & 500   & 28.60 & 21.60 \\
    S1    & 750   & 60.00 & 82.05 & S9    & 750   & 53.65 & 38.85 \\
    S1    & 1000  & 37.10 & 68.15 & S9    & 1000  & 71.35 & 55.25 \\
    S1    & 1250  & 21.55 & 54.15 & S9    & 1250  & 82.70 & 67.55 \\
    S1    & 1500  & 12.70 & 42.35 & S9    & 1500  & 88.55 & 75.60 \\
    S2    & 250   & 73.65 & 78.45 & S10   & 250   & 24.20 & 20.45 \\
    S2    & 500   & 62.30 & 74.65 & S10   & 500   & 52.00 & 43.35 \\
    S2    & 750   & 38.30 & 59.90 & S10   & 750   & 73.10 & 63.95 \\
    S2    & 1000  & 20.80 & 42.50 & S10   & 1000  & 84.00 & 77.60 \\
    S2    & 1250  & 12.70 & 30.20 & S10   & 1250  & 89.75 & 86.30 \\
    S2    & 1500  & 8.60  & 22.15 & S10   & 1500  & 92.65 & 90.45 \\
    S3    & 250   & 16.00 & 12.45 & S11   & 250   & 80.75 & 80.70 \\
    S3    & 500   & 53.25 & 38.00 & S11   & 500   & 93.25 & 93.55 \\
    S3    & 750   & 84.35 & 68.20 & S11   & 750   & 96.50 & 96.55 \\
    S3    & 1000  & 95.80 & 86.00 & S11   & 1000  & 97.10 & 97.50 \\
    S3    & 1250  & 99.00 & 93.95 & S11   & 1250  & 97.15 & 97.60 \\
    S3    & 1500  & 99.75 & 97.30 & S11   & 1500  & 97.45 & 97.95 \\
    S4    & 250   & 43.40 & 36.75 & S12   & 250   & 86.80 & 86.95 \\
    S4    & 500   & 82.10 & 72.75 & S12   & 500   & 94.85 & 95.90 \\
    S4    & 750   & 95.75 & 91.50 & S12   & 750   & 97.00 & 97.35 \\
    S4    & 1000  & 98.60 & 97.35 & S12   & 1000  & 97.25 & 98.30 \\
    S4    & 1250  & 99.45 & 99.00 & S12   & 1250  & 97.45 & 98.85 \\
    S4    & 1500  & 99.95 & 99.45 & S12   & 1500  & 97.45 & 98.80 \\
    S5    & 250   & 67.60 & 70.15 & S13   & 250   & 96.95 & 97.20 \\
    S5    & 500   & 76.95 & 81.45 & S13   & 500   & 99.35 & 99.50 \\
    S5    & 750   & 76.10 & 82.80 & S13   & 750   & 99.55 & 99.75 \\
    S5    & 1000  & 70.50 & 80.50 & S13   & 1000  & 99.40 & 99.70 \\
    S5    & 1250  & 64.45 & 77.10 & S13   & 1250  & 99.30 & 99.60 \\
    S5    & 1500  & 59.40 & 74.25 & S13   & 1500  & 99.05 & 99.55 \\
    S6    & 250   & 68.80 & 70.95 & S14   & 250   & 96.85 & 97.90 \\
    S6    & 500   & 73.60 & 78.60 & S14   & 500   & 98.95 & 99.60 \\
    S6    & 750   & 69.50 & 77.20 & S14   & 750   & 98.75 & 99.45 \\
    S6    & 1000  & 64.70 & 74.95 & S14   & 1000  & 98.55 & 99.35 \\
    S6    & 1250  & 59.25 & 72.85 & S14   & 1250  & 98.00 & 99.35 \\
    S6    & 1500  & 56.55 & 70.75 & S14   & 1500  & 97.85 & 99.20 \\
    S7    & 250   & 44.35 & 42.50 & S15   & 250   & 30.35 & 29.90 \\
    S7    & 500   & 68.35 & 64.15 & S15   & 500   & 42.20 & 42.75 \\
    S7    & 750   & 81.75 & 77.35 & S15   & 750   & 48.45 & 49.35 \\
    S7    & 1000  & 88.95 & 84.85 & S15   & 1000  & 49.30 & 51.55 \\
    S7    & 1250  & 92.45 & 88.35 & S15   & 1250  & 49.90 & 52.20 \\
    S7    & 1500  & 94.55 & 91.40 & S15   & 1500  & 49.85 & 52.30 \\
    S8    & 250   & 58.30 & 56.35 & S16   & 250   & 35.25 & 35.45 \\
    S8    & 500   & 80.50 & 78.30 & S16   & 500   & 46.50 & 47.65 \\
    S8    & 750   & 87.95 & 86.40 & S16   & 750   & 48.55 & 51.85 \\
    S8    & 1000  & 91.45 & 91.40 & S16   & 1000  & 49.45 & 52.95 \\
    S8    & 1250  & 93.55 & 94.05 & S16   & 1250  & 48.85 & 55.00 \\
    S8    & 1500  & 94.85 & 95.50 & S16   & 1500  & 49.15 & 55.65 \\
\end{longtable}

\subsection{Influence of caliper}
In simulation scenarios S9-S16, the performance of the FS-AT test is generally comparable under different caliper choices, including the combined caliper strategy, and those differences observed are as expected. In simulation scenario S9, increased caliper values and adopting the combined caliper strategy achieve lower power for addressing the time-to-hospitalization endpoint with spurious negative “treatment effects”. In simulation scenarios S15\&S16, on the contrary, higher power is achieved by them due to the easier-to-observe treatment effects on the time-to-hospitalization endpoint.

\begin{table}[htbp]
  \centering
  \caption{Empirical power (\%) obtained by the FS-AT test with different calipers $c$ in simulation scenarios S9-S16}
    \begin{tabular}{lrrrrr}
    \toprule
    Scenario & \multicolumn{1}{l}{FU} & \multicolumn{1}{l}{Caliper-10\%} & \multicolumn{1}{l}{Caliper-20\%} & \multicolumn{1}{l}{Caliper-40\%} & \multicolumn{1}{l}{Combined Caliper} \\
    \midrule
    S9    & 500   & 23.20 & 21.60 & 21.60 & 17.80 \\
    S9    & 1000  & 62.70 & 55.25 & 51.30 & 43.60 \\
    S9    & 1500  & 81.70 & 75.60 & 70.25 & 62.10 \\
    S10   & 500   & 47.30 & 43.35 & 39.95 & 37.40 \\
    S10   & 1000  & 81.30 & 77.60 & 72.15 & 70.10 \\
    S10   & 1500  & 91.60 & 90.45 & 86.65 & 85.80 \\
    S11   & 500   & 93.35 & 93.55 & 93.65 & 93.60 \\
    S11   & 1000  & 97.45 & 97.50 & 97.60 & 97.50 \\
    S11   & 1500  & 97.85 & 97.95 & 98.05 & 98.05 \\
    S12   & 500   & 95.45 & 95.90 & 96.20 & 96.20 \\
    S12   & 1000  & 97.90 & 98.30 & 98.55 & 98.60 \\
    S12   & 1500  & 98.25 & 98.80 & 99.20 & 99.30 \\
    S13   & 500   & 99.50 & 99.50 & 99.50 & 99.50 \\
    S13   & 1000  & 99.65 & 99.70 & 99.70 & 99.75 \\
    S13   & 1500  & 99.50 & 99.55 & 99.60 & 99.70 \\
    S14   & 500   & 99.30 & 99.60 & 99.60 & 99.75 \\
    S14   & 1000  & 99.15 & 99.35 & 99.75 & 99.80 \\
    S14   & 1500  & 98.60 & 99.20 & 99.75 & 99.75 \\
    S15   & 500   & 42.65 & 42.75 & 42.55 & 43.00 \\
    S15   & 1000  & 50.35 & 51.55 & 51.35 & 51.40 \\
    S15   & 1500  & 51.35 & 52.30 & 53.10 & 53.75 \\
    S16   & 500   & 47.20 & 47.65 & 49.45 & 48.90 \\
    S16   & 1000  & 50.95 & 52.95 & 55.80 & 55.75 \\
    S16   & 1500  & 52.30 & 55.65 & 58.85 & 60.10 \\
    \bottomrule
    \end{tabular}%
  \label{apx tab:caliper}%
\end{table}%

\subsection{Influence of weight}
In simulation scenarios S9-S16, the performance of the FS-AT test is generally comparable under different weight choices, including the combined caliper strategy, and those differences observed are as expected. In simulation scenario S9, increased weights achieve lower power for addressing the time-to-hospitalization endpoint with spurious negative “treatment effects”. In simulation scenarios S15\&S16, on the contrary, higher power is achieved by increasing the weight due to the easier-to-observe treatment effects on the time-to-hospitalization endpoint.

\begin{table}[htbp]
  \centering
  \caption{Empirical power (\%) obtained by the FS-AT test with different weights $w$ in simulation scenarios S9-S16}
    \begin{tabular}{lrrrrr}
    \toprule
    Scenario & \multicolumn{1}{l}{FU} & \multicolumn{1}{l}{FS-AT(0.1)} & \multicolumn{1}{l}{FS-AT(0.3)} & \multicolumn{1}{l}{FS-AT(0.5)} & \multicolumn{1}{l}{FS-AT(1.0)} \\
    \midrule
    S9    & 500   & 28.60 & 26.00 & 23.65 & 21.60 \\
    S9    & 1000  & 71.30 & 64.75 & 60.75 & 55.25 \\
    S9    & 1500  & 88.05 & 83.20 & 79.90 & 75.60 \\
    S10   & 500   & 52.00 & 47.05 & 44.90 & 43.35 \\
    S10   & 1000  & 83.50 & 80.00 & 78.65 & 77.60 \\
    S10   & 1500  & 92.20 & 91.25 & 91.15 & 90.45 \\
    S11   & 500   & 93.25 & 93.50 & 93.65 & 93.55 \\
    S11   & 1000  & 97.10 & 97.50 & 97.40 & 97.50 \\
    S11   & 1500  & 97.35 & 97.70 & 97.95 & 97.95 \\
    S12   & 500   & 94.85 & 95.65 & 95.70 & 95.90 \\
    S12   & 1000  & 97.40 & 98.30 & 98.35 & 98.30 \\
    S12   & 1500  & 98.00 & 98.40 & 98.65 & 98.80 \\
    S13   & 500   & 99.35 & 99.45 & 99.50 & 99.50 \\
    S13   & 1000  & 99.40 & 99.55 & 99.65 & 99.70 \\
    S13   & 1500  & 99.15 & 99.40 & 99.50 & 99.55 \\
    S14   & 500   & 98.95 & 99.25 & 99.50 & 99.60 \\
    S14   & 1000  & 98.70 & 99.20 & 99.30 & 99.35 \\
    S14   & 1500  & 98.35 & 98.75 & 99.15 & 99.20 \\
    S15   & 500   & 42.20 & 42.40 & 42.75 & 42.75 \\
    S15   & 1000  & 49.30 & 51.20 & 51.25 & 51.55 \\
    S15   & 1500  & 50.00 & 51.80 & 52.10 & 52.30 \\
    S16   & 500   & 46.50 & 47.95 & 47.90 & 47.65 \\
    S16   & 1000  & 49.85 & 52.25 & 52.80 & 52.95 \\
    S16   & 1500  & 50.55 & 54.05 & 55.15 & 55.65 \\
    \bottomrule
    \end{tabular}%
  \label{apx tab:weight}%
\end{table}%

\section{
Influence of correlation
}
\label{apx see: correlation}
In this appendix, we provide additional analysis of the influence that the correlation between survival time and time-to-hospitalization endpoints has on the power. The overall NB and WR are decomposed to endpoint-level, i.e., NB and WR for the survival time and time-to-hospitalization endpoints. Because the merging of multiple stages corresponding to one endpoint is non-trivial for WO due to the specific handling of ties, WO is not reported at the endpoint level.
In the FS test, comparison results in stages 1\&2 are attributed to survival time and time-to-hospitalization endpoints correspondingly. In the FS-AT test, stages 1\&3 are attributed to the survival time endpoint, and stages 2\&4 are attributed to the time-to-hospitalization endpoint. 
In simulation scenario S3, both FS-AT and FS tests have their time-to-hospitalization endpoint NB being negative, as opposed to those NB close to 0 in simulation scenario S4.
Similarly, in simulation scenario S3, both FS-AT and FS tests have their time-to-hospitalization endpoint WR less then 1, as opposed to those WR close to 1 in simulation scenario S4.
Such a comparison indicates the existence of spurious negative “treatment effects” introduced by the correlation between survival time and time-to-hospitalization endpoints.

\begin{table}[htbp]
  \centering
  \caption{Endpoint-level decomposition of NB(\%) in simulation scenarios S3\&S4 averaged across 2000 replicates}
  \resizebox{\textwidth}{!}{%
    \begin{tabular}{lrrrrrrr}
    \toprule
    \multicolumn{1}{c}{\multirow{2}[2]{*}{Scenario}} & \multicolumn{1}{c}{\multirow{2}[2]{*}{FU}} & \multicolumn{3}{c}{FS} & \multicolumn{3}{c}{FS-AT} \\
          &       & Overall & Survival Time & Time to Hospitalization & Overall & Survival Time & Time to Hospitalization \\
    \midrule
    S3    & 250   & 2.35  & 4.42  & -2.07 & 2.00  & 4.02  & -2.02 \\
    S3    & 500   & 5.15  & 7.55  & -2.40 & 4.24  & 6.89  & -2.65 \\
    S3    & 750   & 7.58  & 9.72  & -2.14 & 6.19  & 9.00  & -2.81 \\
    S3    & 1000  & 9.52  & 11.26 & -1.74 & 7.78  & 10.58 & -2.80 \\
    S3    & 1250  & 10.98 & 12.34 & -1.36 & 9.04  & 11.76 & -2.72 \\
    S3    & 1500  & 12.07 & 13.10 & -1.04 & 10.00 & 12.65 & -2.65 \\
    S4    & 250   & 4.26  & 4.29  & -0.03 & 3.88  & 3.91  & -0.03 \\
    S4    & 500   & 7.37  & 7.41  & -0.04 & 6.59  & 6.61  & -0.02 \\
    S4    & 750   & 9.58  & 9.62  & -0.04 & 8.65  & 8.69  & -0.04 \\
    S4    & 1000  & 11.12 & 11.14 & -0.03 & 10.23 & 10.25 & -0.02 \\
    S4    & 1250  & 12.22 & 12.23 & -0.01 & 11.40 & 11.42 & -0.02 \\
    S4    & 1500  & 13.01 & 13.01 & -0.01 & 12.30 & 12.31 & -0.01 \\
    \bottomrule
    \end{tabular}%
    }
  \label{apx tab:correlation_NB}%
\end{table}%

\begin{table}[htbp]
  \centering
  \caption{Endpoint-level decomposition of WR in simulation scenarios S3\&S4 averaged across 2000 replicates}
  \resizebox{\textwidth}{!}{%
    \begin{tabular}{lrrrrrrr}
    \toprule
    \multicolumn{1}{c}{\multirow{2}[2]{*}{Scenario}} & \multicolumn{1}{c}{\multirow{2}[2]{*}{FU}} & \multicolumn{3}{c}{FS} & \multicolumn{3}{c}{FS-AT} \\
          &       & \multicolumn{1}{l}{Overall} & \multicolumn{1}{l}{Survival Time} & \multicolumn{1}{l}{Time to Hospitalization} & \multicolumn{1}{l}{Overall} & \multicolumn{1}{l}{Survival Time} & \multicolumn{1}{l}{Time to Hospitalization} \\
    \midrule
    S3    & 250   & 1.07  & 1.36  & 0.90  & 1.06  & 1.39  & 0.91 \\
    S3    & 500   & 1.12  & 1.36  & 0.89  & 1.10  & 1.38  & 0.89 \\
    S3    & 750   & 1.17  & 1.36  & 0.88  & 1.14  & 1.38  & 0.87 \\
    S3    & 1000  & 1.21  & 1.36  & 0.87  & 1.17  & 1.39  & 0.85 \\
    S3    & 1250  & 1.25  & 1.35  & 0.86  & 1.20  & 1.39  & 0.82 \\
    S3    & 1500  & 1.28  & 1.35  & 0.85  & 1.22  & 1.40  & 0.80 \\
    S4    & 250   & 1.12  & 1.35  & 1.00  & 1.11  & 1.35  & 1.00 \\
    S4    & 500   & 1.17  & 1.35  & 1.00  & 1.15  & 1.36  & 1.00 \\
    S4    & 750   & 1.22  & 1.35  & 1.00  & 1.19  & 1.37  & 1.00 \\
    S4    & 1000  & 1.25  & 1.35  & 1.00  & 1.23  & 1.38  & 1.00 \\
    S4    & 1250  & 1.28  & 1.35  & 1.00  & 1.26  & 1.38  & 1.00 \\
    S4    & 1500  & 1.30  & 1.35  & 1.00  & 1.28  & 1.39  & 1.00 \\
    \bottomrule
    \end{tabular}%
  }
  \label{apx tab:correlation}%
\end{table}

\newpage
\section{
Additional details of the case study
}

The number of patients within each stratum in the case study is presented in Table \ref{tab:stratum num}.

\begin{table}[htbp]
  \centering
  \captionsetup{justification=centering}
  \caption{Number of patients within each stratum}
    \begin{tabular}{lrrrr}
    \toprule
          & \multicolumn{4}{c}{Ejection fraction $<$ 0.25} \\
          & \multicolumn{2}{c}{Ischemic } & \multicolumn{2}{c}{Nonischemic} \\
          & \multicolumn{1}{c}{Age $<$ 70} & \multicolumn{1}{c}{Age $\geq$ 70} & \multicolumn{1}{c}{Age $<$ 70} & \multicolumn{1}{c}{Age $\geq$ 70} \\
    \midrule
    Treatment & 231   & 82    & 102   & 53 \\
    Control & 190   & 119   & 136   & 39 \\
    \midrule
          & \multicolumn{4}{c}{Ejection fraction 0.25-0.45} \\
          & \multicolumn{2}{c}{Ischemic } & \multicolumn{2}{c}{Nonischemic} \\
          & \multicolumn{1}{c}{Age $<$ 70} & \multicolumn{1}{c}{Age $\geq$ 70} & \multicolumn{1}{c}{Age $<$ 70} & \multicolumn{1}{c}{Age $\geq$ 70} \\
    \midrule
    Treatment & 286   & 188   & 113   & 61 \\
    Control & 277   & 184   & 100   & 56 \\
    \bottomrule
    \end{tabular}%
  \label{tab:stratum num}%
\end{table}%

\end{document}